\documentclass[preprint,prb]{revtex4-1}

\usepackage[utf8]{inputenc}
\usepackage[english]{babel}
\usepackage{amsmath,amsfonts,amssymb}
\usepackage{graphicx,pstricks,subfig,caption}

\begin{document}

\title{Exploring charge density distribution and electronic properties of hybrid organic-germanium layers}
\author{Fl\'avio Bento de Oliveira}
\affiliation{Universidade Federal de Goi\'as, Institute of Physics, Campus Samambaia, 74690-900 Goi\^ania, Goi\'as, Brazil}
\author{Erika Nascimento Lima}
\affiliation{Universidade Federal de Rondon\'opolis, ICEN, Rondon\'opolis, Mato Grosso, Brazil}
\author{Andreia Luisa da Rosa}
\affiliation{Universidade Federal de Goi\'as, Institute of Physics, Campus Samambaia, 74690-900 Goi\^ania, Goi\'as, Brazil}
\affiliation{Bremen Center for Computational Materials Science, University of Bremen, Am Fallturm 1, 28359 Bremen, Germany}
\author{Mauricio Chagas da Silva}
\affiliation{Bremen Center for Computational Materials Science, University of Bremen, Am Fallturm 1, 28359 Bremen, Germany}
\affiliation{Max Planck Institute for the Structure and Dynamics of Matter; Luruper Chaussee 149, Geb. 99 (CFEL), 22761 Hamburg, Germany}
\affiliation{Université de Lorraine, CNRS, LPCT, F-54000 Nancy, France}
\author{Thomas Frauenheim}
\affiliation{Bremen Center for Computational Materials Science, University of Bremen, Am Fallturm 1, 28359 Bremen, Germany}
\affiliation{Computational Science Research Center, No.10 East Xibeiwang Road, Beijing 100193 and Computational Science and Applied Research Institute Shenzhen, China.}

\begin{abstract}

  Band gap tuning and dielectric properties of small organic ligands
  adsorbed on bidimensional germanium monolayers (germanene) have been
  investigated using first-principles calculations. We show that the
  adsorption of these small groups retains the initially stable
  free-standing pristine buckled germanium nanostructures. Charge
  density and chemical bonding analysis show that the ligands are
  chemisorbed on the germanium layers. Finally we demosntrate that the
  dielectric properties of bare and ligand adsorbed germanene have a
  large anisotropy. Our findings of a finite gap shows open a path for
  rational design of nanostructures with possible applications in
  biosensors and solar cells.

\end{abstract}

%\pacs{PACS numbers: 75.60.Ej, 64.60.Ak}

\maketitle

\section{Introduction}

The study of two-dimensional materials has increasing interest after
the discovery of graphene whose property differed surprisingly from
its three-dimensionl form, graphite\,\cite{Novoselov:04,Geim:07}. These
materials have electronic properties of great interest for
technological applications. Graphene in its honeycomb structure has a zero gap, with
conducion and valence bands being degenerate at K and K points,
forming cones. However, difficulties in carrying out and adjusting a
reasonable sizeable band gap in graphene is attracting increasing
interest in other two-dimensional materials.

Recent investigations suggest that germanene react rapidly with the
environment\,\cite{JiangNT:2014}. This may affect not only their
electronic structure, but also their reactivity, dielectric and
optical properties. Therefore one may search for ways of tuning the
electronic properties of these two-dimensional structures. A promising
route waas demonstrated using adsorption of organic molecules or functional groups
\,\cite{ProgressReports}.  It has been
shown that modified germanene were stable hybrid materials and with
tunable optical properties\,\cite{JiangNT:2014}. Although bare
germanium layers have been extensively investigated, functionalized
layers with organic molecules has received less attention. Theoretical
investigations by Kou and co-authors proposed that adsorption of ${\rm
  -CH_2OCH_3}$ groups on germanene could lead to ferroelectric
properties\,\cite{Kou2018a}.  Furthermore it has been suggested that
hydrogen adsorbed layers become topological insulators when external strain is
applied\,\cite{Rivelino2015}. 

In this work we have investigated the electronic and dielectric
properties of germanene with small organic groups using
first-principles calculations. We show that the some of the adsorbed
structures possess a sizeable band gap.  We claim that the stability
of these structure is a combined effect of both in-plane strain
induced by the adsorption of organic groups and ligand-ligand
interactions. Finally we show that the dielectric properties of this
material have a large anisotropy. 

\section{Methodology}

We employ density functional theory\,\cite{Hohenberg:64,Kohn:65}
within the generalized gradient approximation\,\cite{Perdew:96} and
the projected augmented wave method
(PAW)\,\cite{Bloechel:94,Kresse:99} as implemented in the VASP
code\,\cite{Kresse:99} to investigate the electronic
structure germanene hybrid layers. A ($1\times$1) supercell containing
two germanium atoms was employed.  Forces on atoms were converged
until 10$^{-4}$\,eV/{\AA}. A vacuum region of 10 {\AA} has been found
to be sufficient to avoid spurious interactions between germanene
layers in neighbouring cells. A (10$\times$10$\times$1) {\bf k}-point
sampling has been used in the calculations of all investigated systems
with an energy cutoff of 500 eV. The calculation of the dielectric
function was performed using the GW method\,\cite{Shishkin:07}. A
(8$\times$8$\times$1) {\bf k}-points mesh has been employed. The
electron localization function visualization was obtained using the
VESTA software \cite{Momma2011} and the critical point analysis was provided 
by the post-process of the total electronic density using Critic2 
software\cite{Otero-de-la-Roza2009,Otero-De-La-Roza2014}.

\section{Results and Discussions}

\subsection{Structural properties}

\begin{figure}[htpb!]
  \includegraphics[width=7cm]{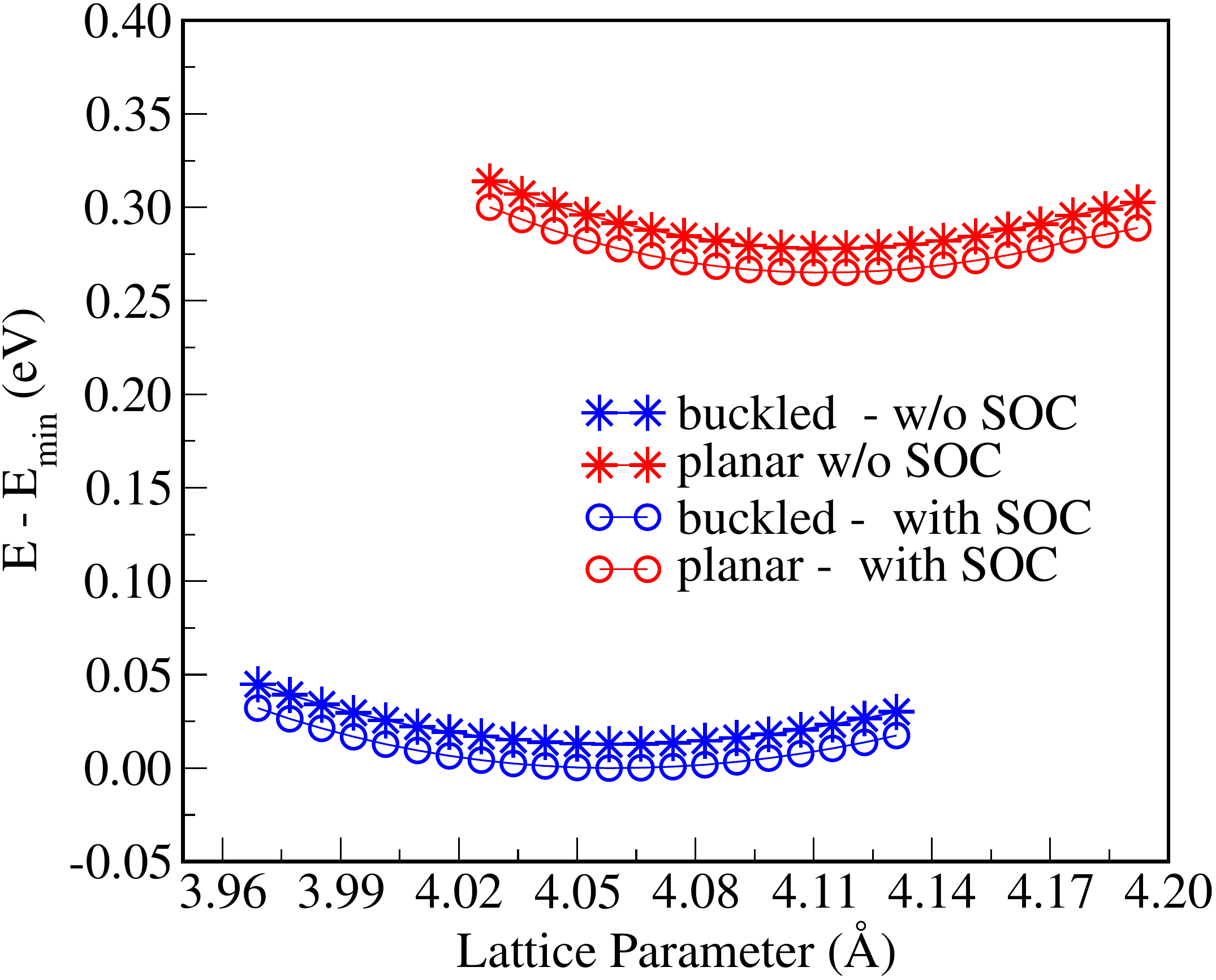}
  \caption{\label{fig:stability}(color online) Total energy of buckled and planar
     germanium layers with and without spin-orbit coupling. The zero of energy is set 
     to the lowest energy of the buckled structure calculated within GGA.}
\end{figure}

\begin{figure}[htpb!]
  \centering
  \subfloat[]{\includegraphics[width=3.0cm]{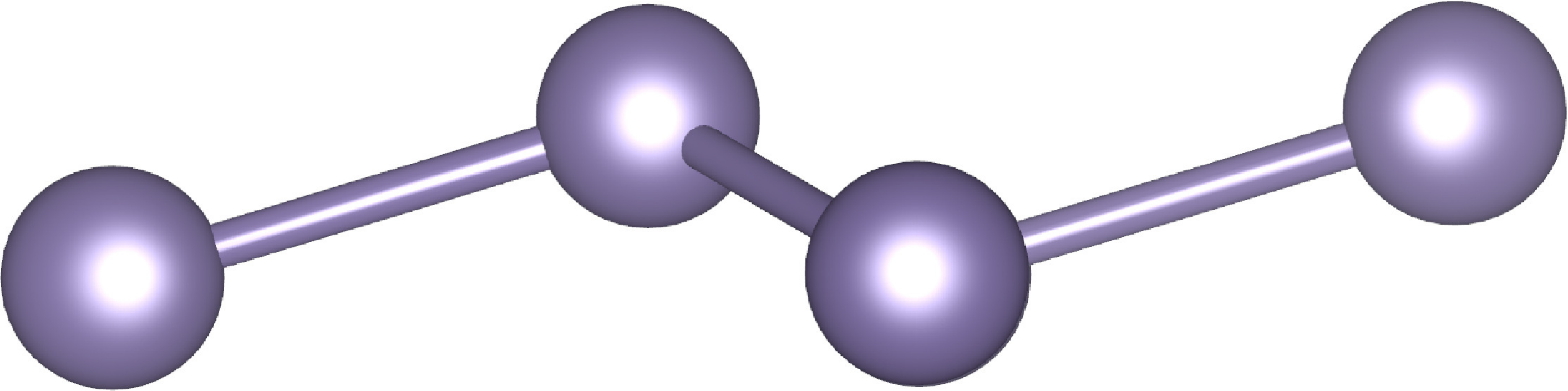}}
  \subfloat[]{\includegraphics[width=3.0cm]{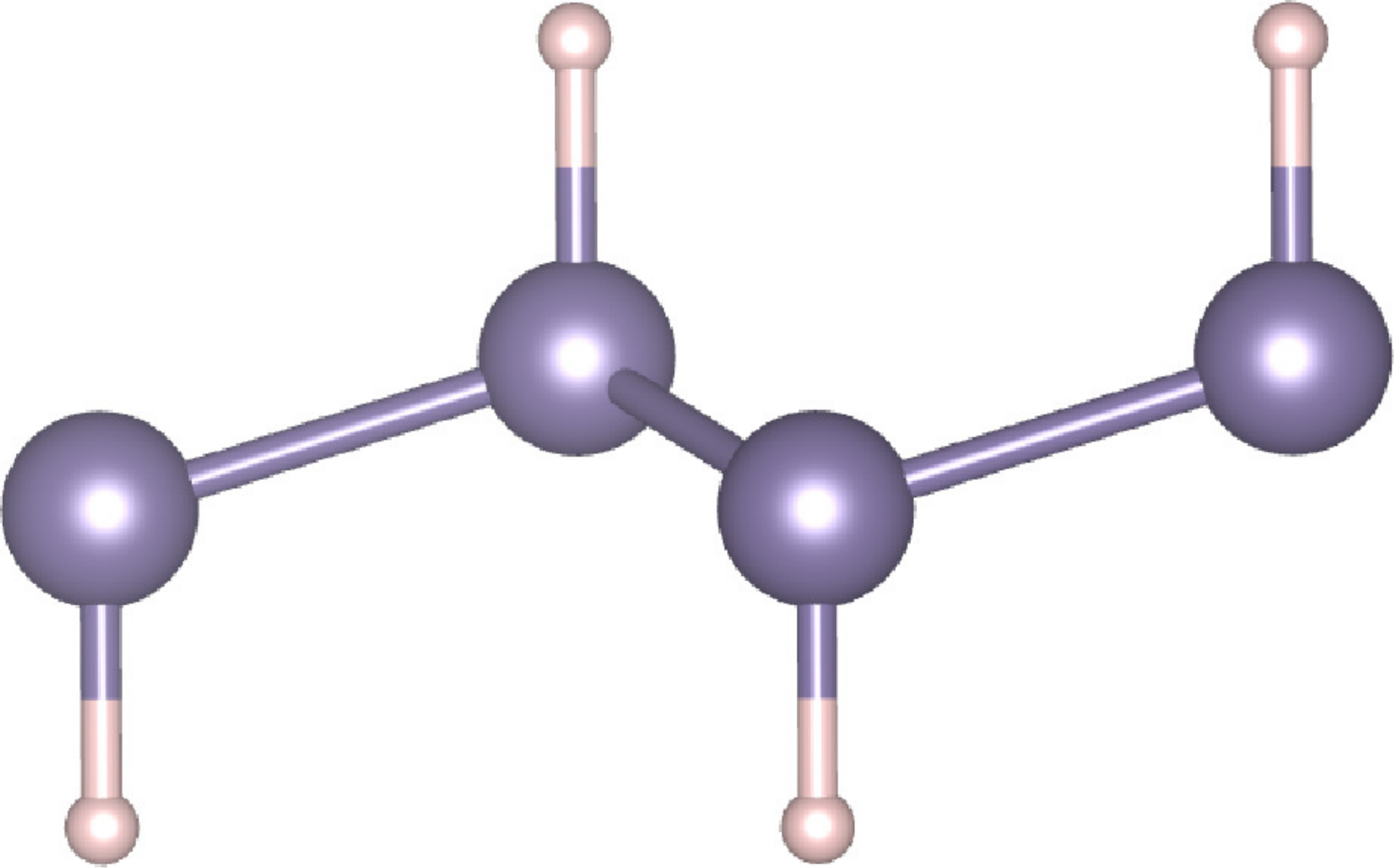}}
  \subfloat[]{\includegraphics[width=3.0cm]{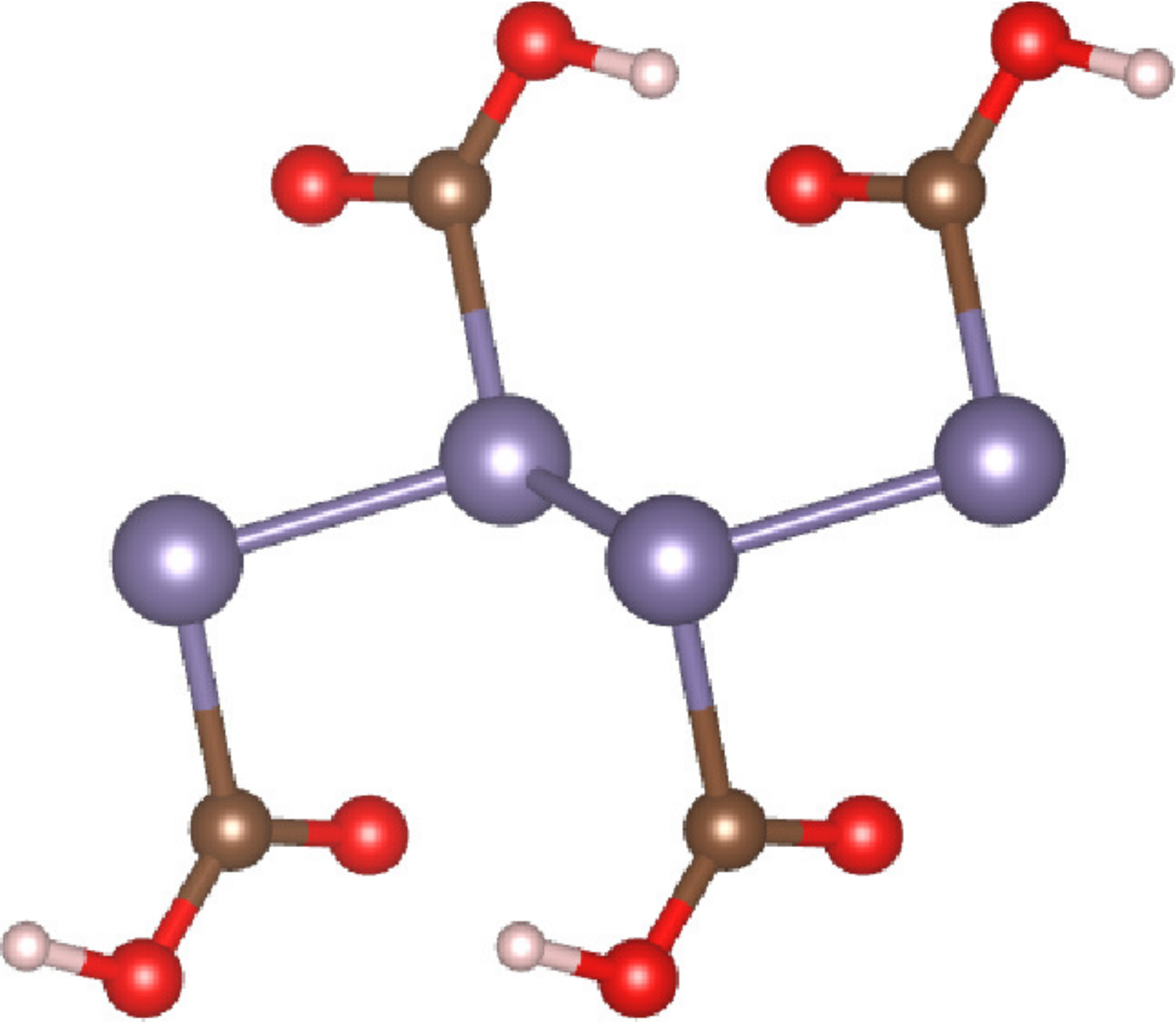}}
  \\
  \subfloat[]{\includegraphics[width=3.0cm]{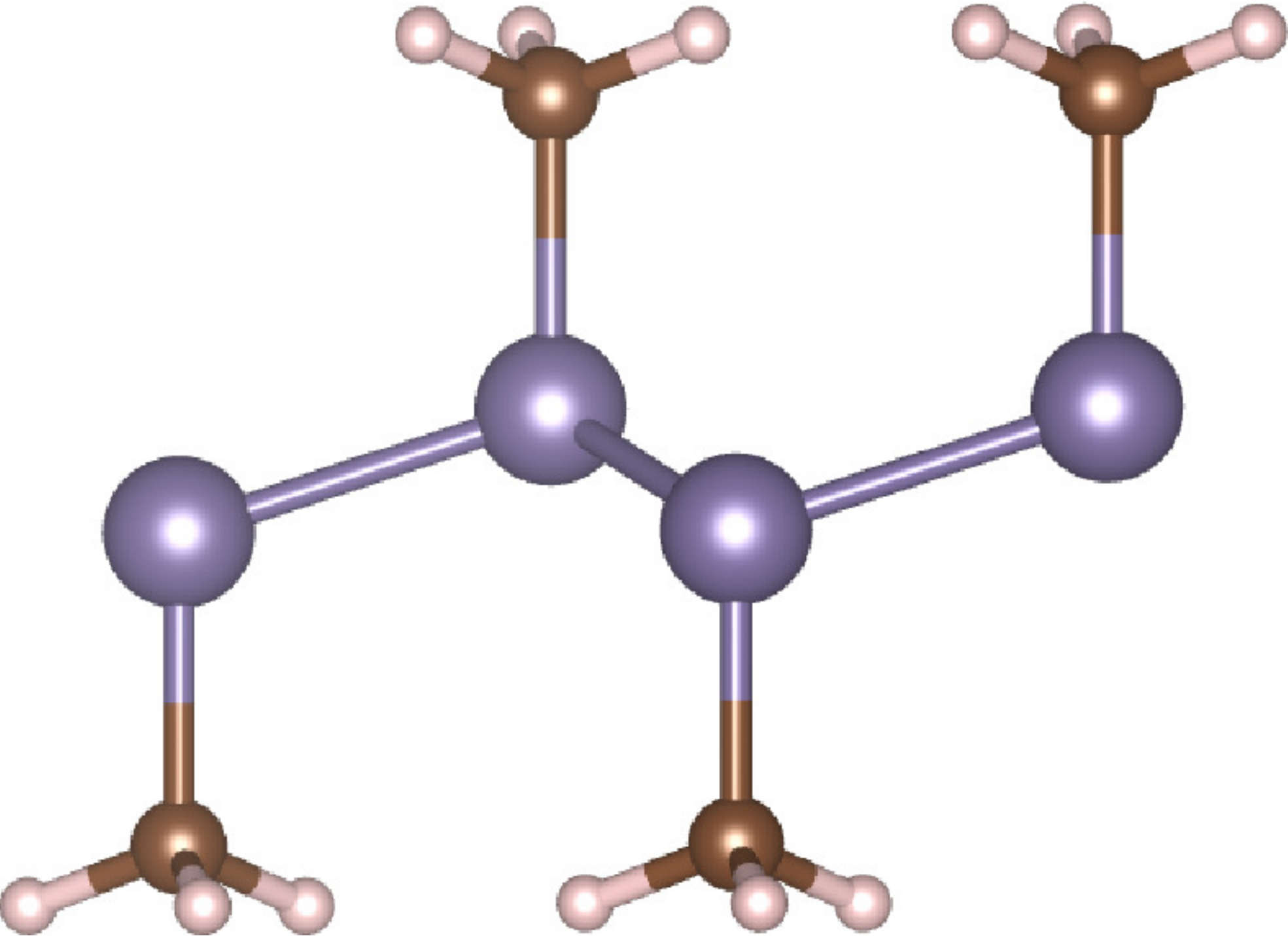}}
  \subfloat[]{\includegraphics[width=3.0cm]{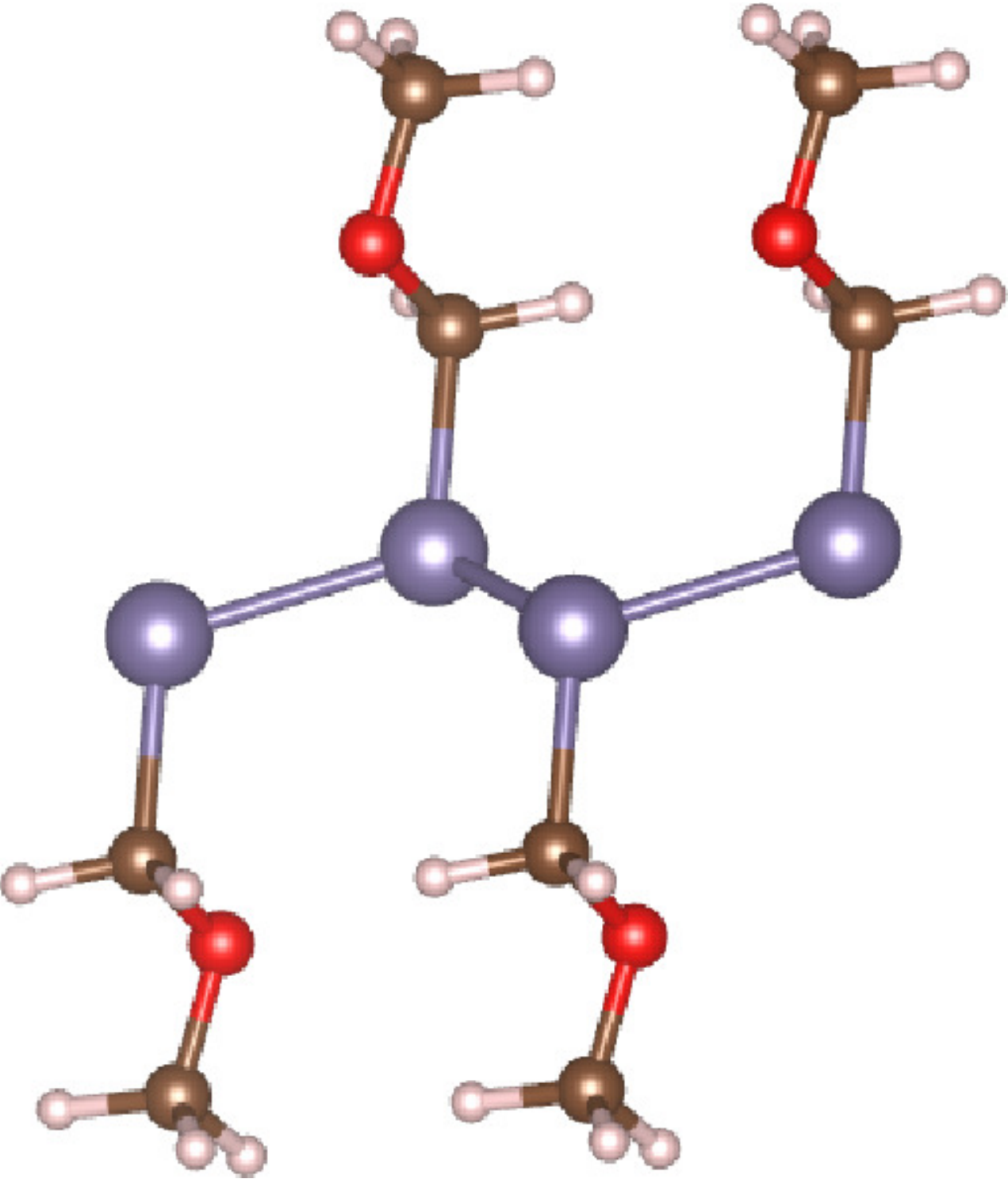}}
  \subfloat[]{\includegraphics[width=3.0cm]{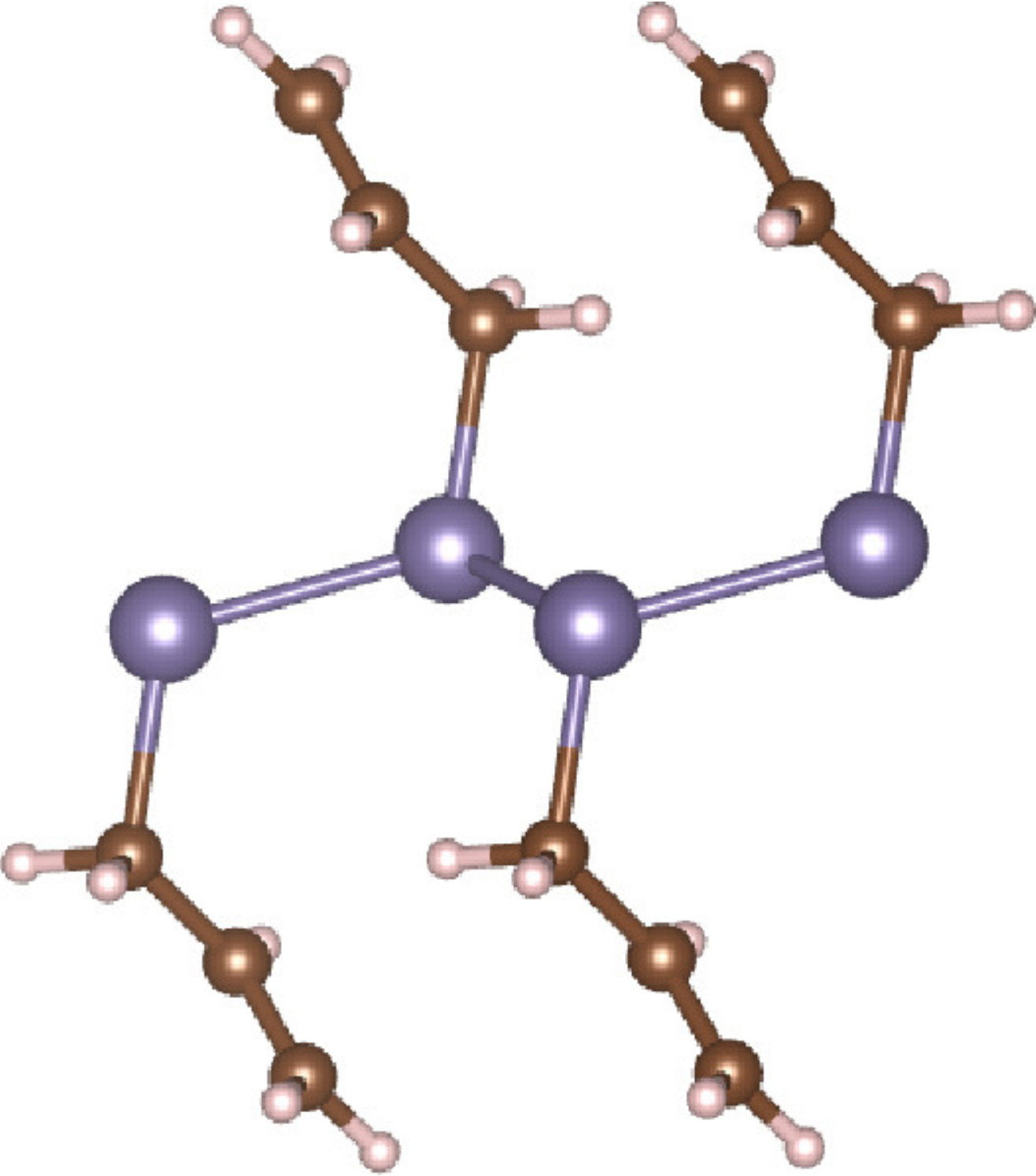}}
    \caption{\label{fig:structures}(color online) Side view of relaxed germanium layers
    adsorbed with organic ligands at 1 ML regime: a) buckled bare, b) hydrogen, c) -COOH, 
    d) -CH$_3$, e)  CH$_2$OCH$_3$ and f) CH$_2$CHCH$_2$. Red, brown, blue, white and 
    magenta are oxygen, carbon, hydrogen and germanium atoms, respectively.}
\end{figure}

The structural stability of the buckled and planar structure has been
investigated by varying the in-plane lattice parameter $a$ and fully
relaxing all the atoms in the unit cell. The relaxed bare buckled
structure is shown in Fig.\,\ref{fig:structures}(a). The inclusion of
spin-orbit coupling does not change the lattice parameter, but it
lowers the total energies of the bare structure as shown in
Fig.\ref{fig:stability}.  The optimized in-plane lattice parameter for
buckled (planar) germanene is $a$ = 4.05{\AA} (4.11{\AA}). The buckled
structure is 0.2\,eV energetically more stable than the
planar one. The Ge-Ge distance is 2.43\,{\AA}, as shown in Table
\ref{table:structure}. This result is in good agreement with previous
calculations\,\cite{Ciraci2019,Kou2018a}.

\begin{table}[htpb!]
\begin{center}
\caption{\label{table:structure} Lattice parameter $a$, Ge-Ge distance, Ge-ligand distance, buckling of germanium modified structures calculated within GGA.}
\begin{tabular*}{1.0\textwidth}{@{\extracolsep{\fill}}lccccc}
\hline
\hline
                 &      $a$    & Ge-Ge    &   Ge-C   & Ge-H & buckling  \\ 
\hline
bare             &    4.05   &   2.43    &          &       &  0.69   \\
-H               &    4.08   &   2.47    &          & 1.57  &  0.74   \\
-COOH            &    4.20   &   2.52    &   2.08   &       &  0.74   \\
-CH3             &    4.11   &   2.50    &   2.00   &       &  0.78  \\
-CH$_2$OCH$_3$   &    4.24   &   2.55    &   2.02   &       &  0.71   \\
-CH$_2$CHCH$_2$  &    4.40   &   2.65    &   2.05   &       &  0.73  \\
\hline
\hline
\end{tabular*}
\end{center}
\end{table}

\clearpage

\clearpage

We have investigated the following ligands: -H, -CH$_3$, -COOH,
-CH$_2$CHCH$_2$ and -CH$_2$OCH$_3$ groups, since these groups were
sinthesized experimentally\,\cite{JiangNT:2014} The relaxed bare and
funcionalized germanene are shown in Fig.\,\ref{fig:structures}. We
have considered structures with ligands adsorbing on top positions of
germanium atoms on both sides of the germanium surface. Although we
cannot rule out that smaller coverages may be present, we should point
out that higher coverages yield to more stable structures in
germanene\,\cite{JiangNT:2014}. Upon adsorption of ligands, the
initially buckled geometry of germanene has a slight different
buckling compared to the bare layers, as shown in
Table\,\ref{table:structure}. This is rather different from what has
been found for similar groups on bismuthene which become planar upon
adsoprtion of ligands\,\cite{JPCC2020,Kou2018a,submitted}. As a general feature, the
intralayer lattice parameter $a$ and consequently Ge-Ge distance
increases as the ligand size increases. This means that the
ligand-ligand interaction plays an important role on the stabilization
of the hybrid structures.

The in-plane lattice parameter of H-Ge shown in
Fig. \ref{fig:structures} (b) is 4.08 {\AA} with Ge-Ge bond length of
2.47({\AA}). The Ge-COOH structure is shown in
Fig. \ref{fig:structures} (c) and has $a$ = 4.20 {\AA} and Ge-Ge
distance of 2.52 {\AA}. The ${\rm Ge-CH_3}$ structure shown in
Fig. \ref{fig:structures} (d) has $a$ = 4.11{\AA} and Ge-Ge distance of
2.50. This is similar to other small groups due to similar van der
Waals radii. The in-plane lattice constant is somewhat larger for
${\rm Ge-CH_2CHCH_2}$ shown in Fig.\,\ref{fig:structures} (e) (4.21 {\AA})
and ${\rm Ge-CH_2OCH_3}$ seen in Fig.\,\ref{fig:structures} (f)
(4.40{\AA}). In particular, the ligands in the later two structures
assume a tilted configuration. One can conclude that the not only the
ligand size but also its character is important to
determine the bond strength.

\section{Chemical bonding analysis}

\begin{figure}[htpb!]
  \centering
  \subfloat[]{\includegraphics[width=3cm]{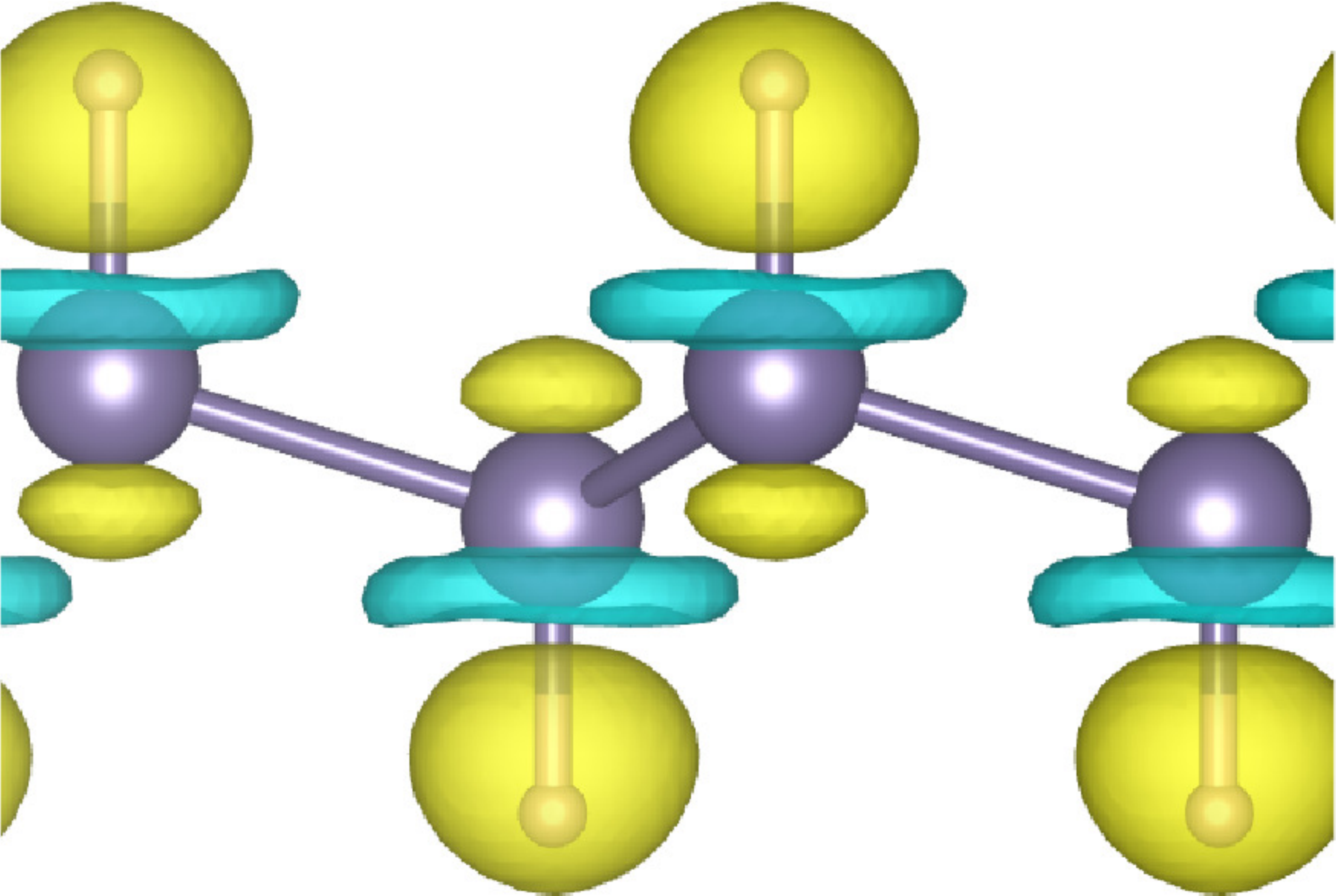}}
  \subfloat[]{\includegraphics[width=3cm]{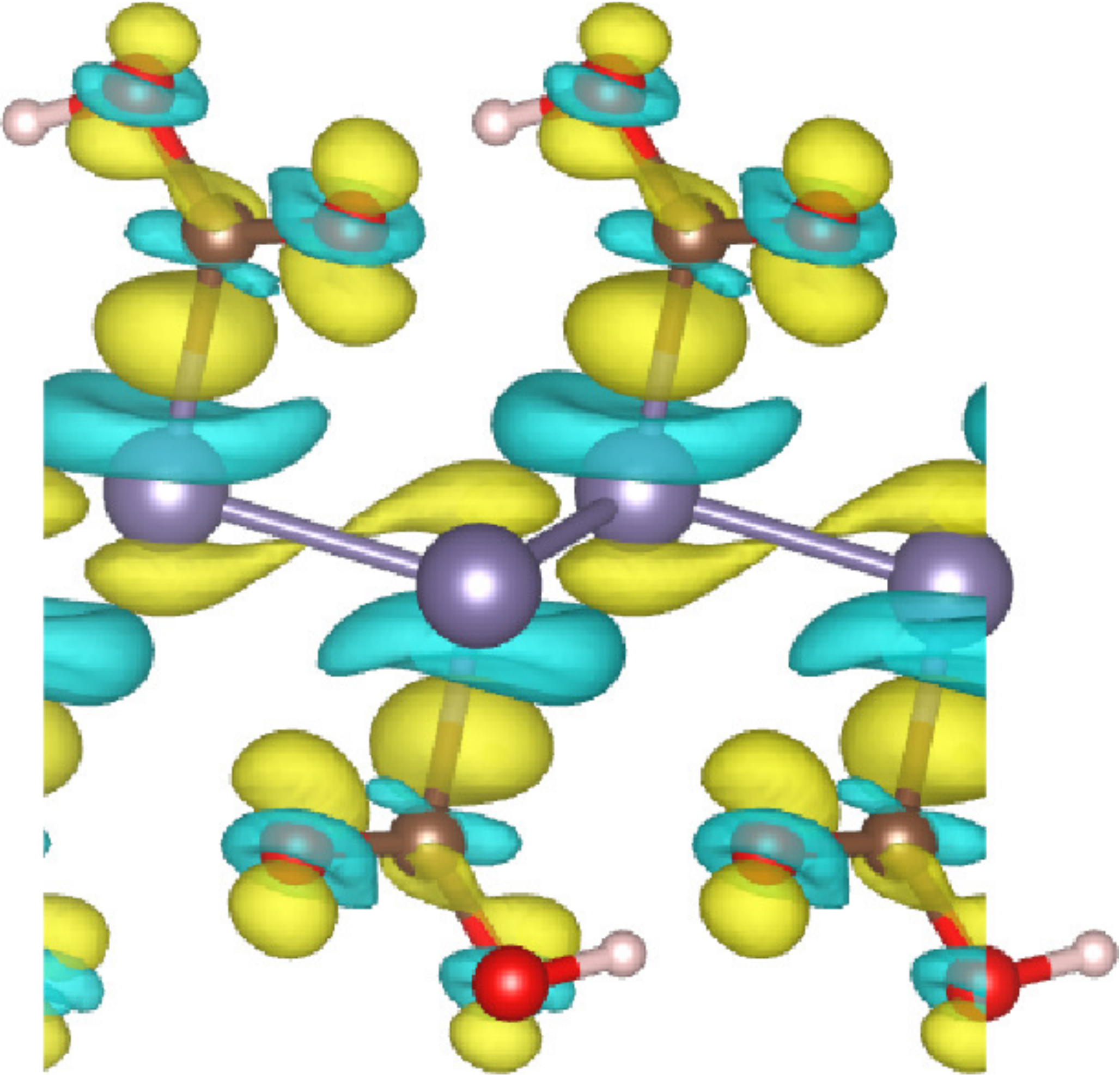}}
  \subfloat[]{\includegraphics[width=3cm]{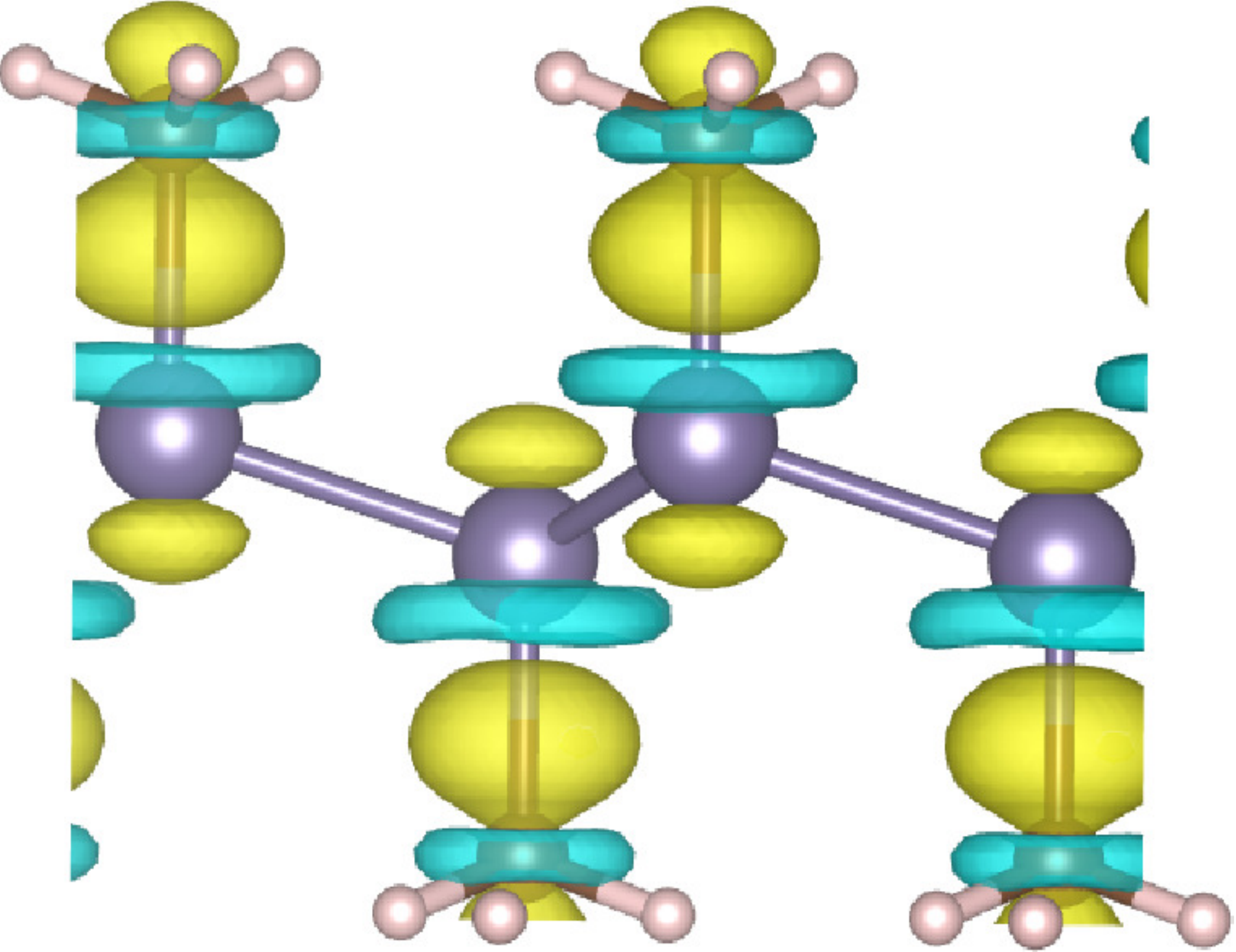}}
  \\
  \subfloat[]{\includegraphics[width=3.5cm]{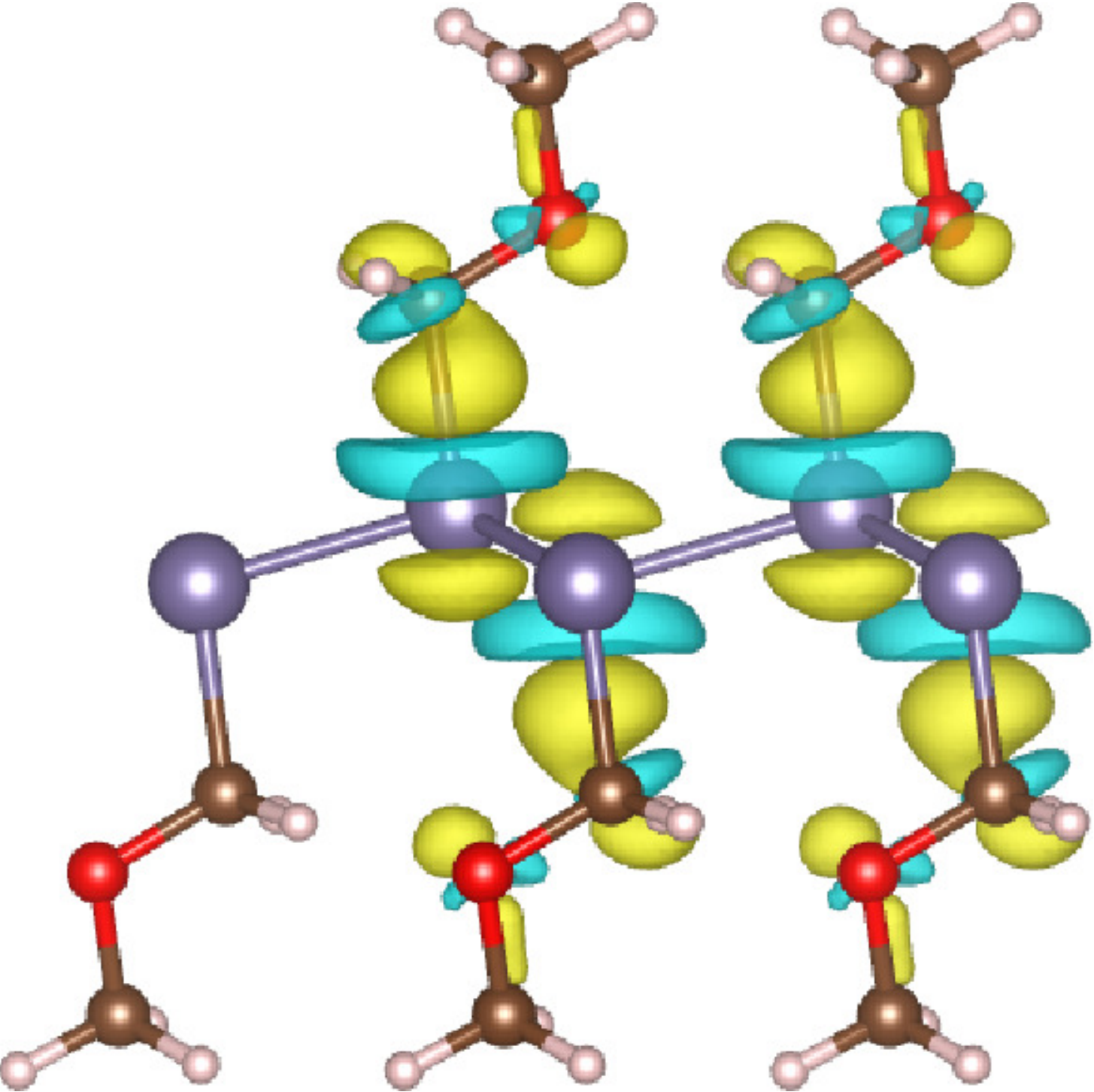}}
  \subfloat[]{\includegraphics[width=3.5cm]{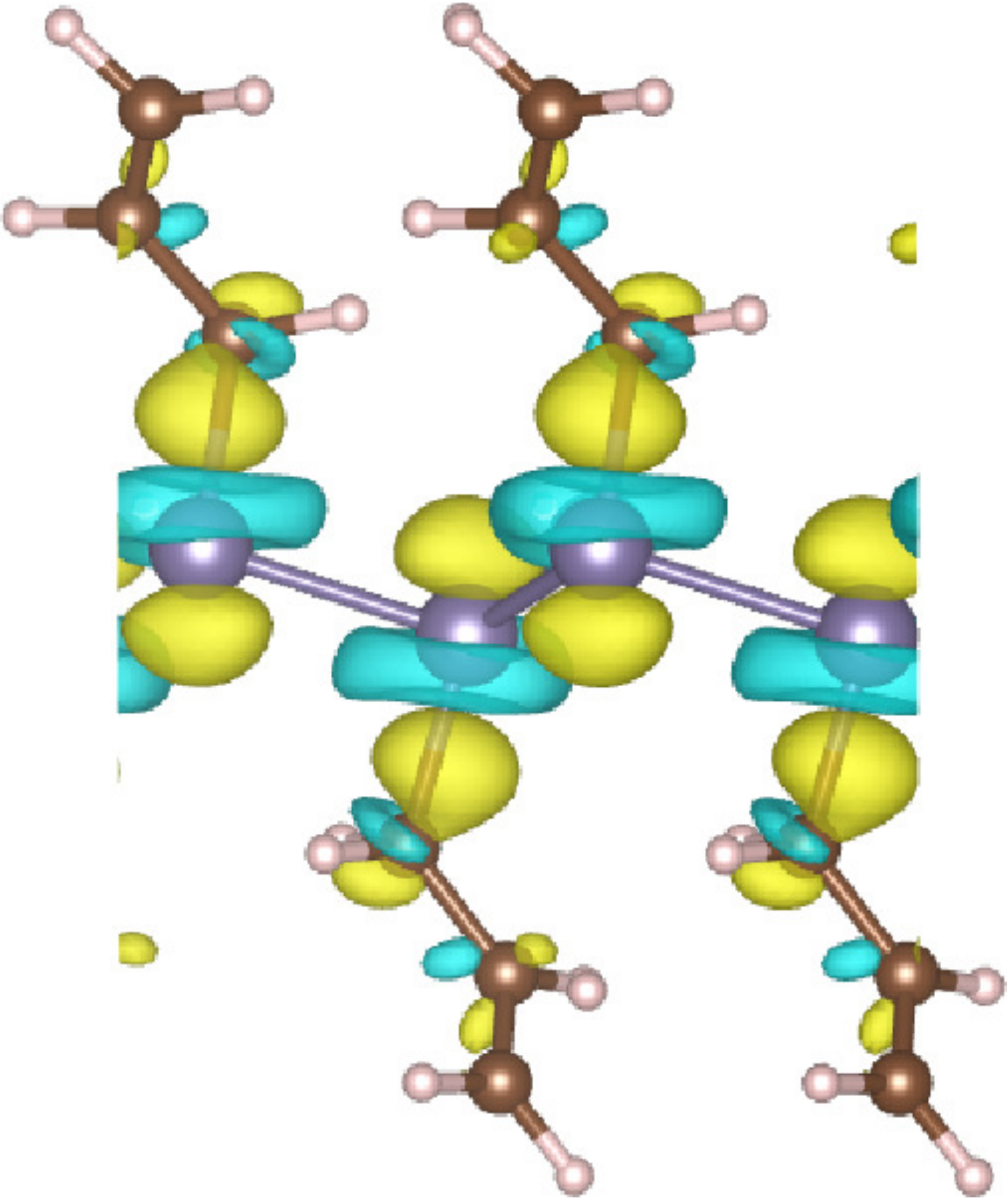}}
  \caption{\label{fig:chargediff}Charge density difference of modified
  germanene structures with a) -H, b) -COOH, c) -CH$_3$, d)
  CH$_2$OCH$_3$ and e) CH$_2$CHCH$_2$. Blue indicates a loss of electrons
  whereas yellow indicates accumulation of electrons. Isosurface
  levels are set to 0.002 $e$/\AA$^3$.}
\end{figure}

The nature of the chemical bonds in the hybrid systems was
investigated by calculating the charge density difference between the
electronic densities of the adsorbed germanium layers and their
constituent systems. The electronic density difference shown in
Fig.\,\ref{fig:chargediff} is given by ${\rm \Delta\rho = \rho^{Ge-X}
  -\rho^{Ge} - \rho^{-X}}$, where ${\rm \rho^{Ge-X}}$ is the
electronic density of the hybrid Ge-X layers, $\rho^{Ge}$ and
${\rho^{-X}}$ are the charge densities of the germanium layers and the
ligand, respectively, calculated at fixed atomic positions after
structure optimization. As a general feature, charge accumulation on
the ligand (yellow) and charge withdrawal (blue) close to the
germanium atom is seen.

\clearpage

Ge-H hybrid layers shown in Fig.\,\ref{fig:chargediff}(a) have a
larger electron density at hydrogen atoms and a withdraw of electrons
at germanium sites. Ge-COOH also shows accumulation of charge at the
ligand, mainly on oxygen and carbon atoms, while charge is withdrawn
at germanium sites as shown in Fig.\,\ref{fig:chargediff}(b). The
complex Ge-CH$_3$ shown in shows that electronic charge is accumulated
the ligand upon adsorption on the germanene. Furthremore, between C
and Ge bonds there is an excess of electrons as it can be seen in
Fig.\,\ref{fig:chargediff}(c) . The Ge-${\rm CH_2OCH_3}$ hybrid also
shows accumulation of charge at the ligand, mainly on the oxygen and
carbon atoms, while charge is depleted at the germanium site, as shown
in Fig.\,\ref{fig:chargediff}(d). The complex Ge-CH$_2$CHCH$_2$ also
shows accumulation of electrons on the ligand oxygen and carbon atoms,
while regions close to the germanium are depleted of electrons. At the
germanium site and also between C and Ge bonds there is an excess of
electrons as seen in Fig.\,\ref{fig:chargediff}(e).  One can therefore
conclude that due to the charge accumulation/withdraw in the modified
germanium layers, the reactivity of the whole system changes. This
could be quite useful for further docking of organic or biomolecules.

\begin{figure}[hbtp!]
  \centering
  \subfloat[bare]{\includegraphics[width=3cm]{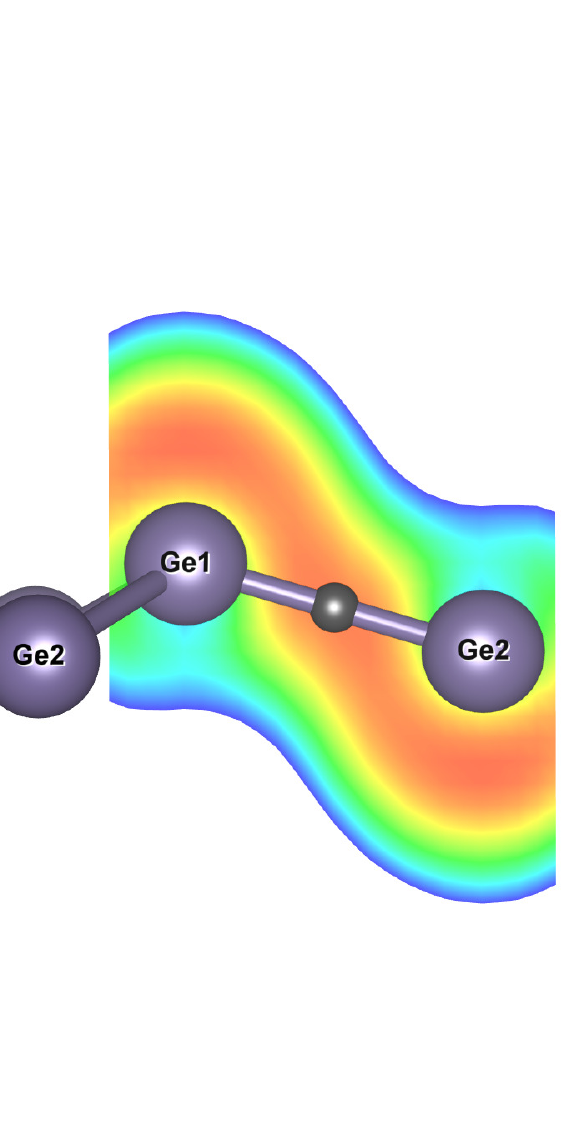}}
  \subfloat[-H]{\includegraphics[width=2.5cm]{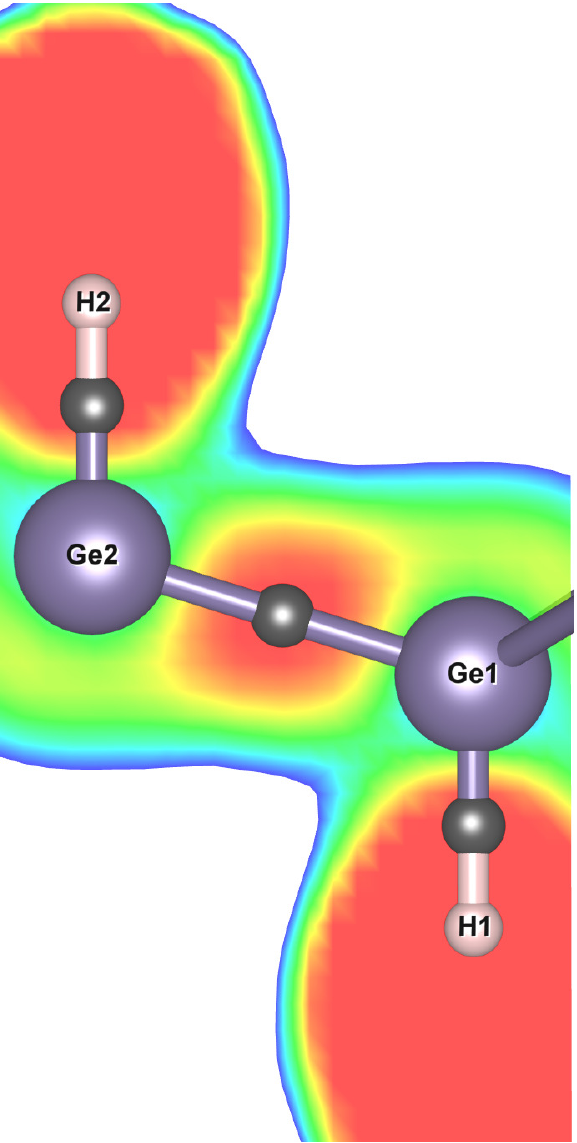}}
  \subfloat[-CH$_3$]{\includegraphics[width=3cm]{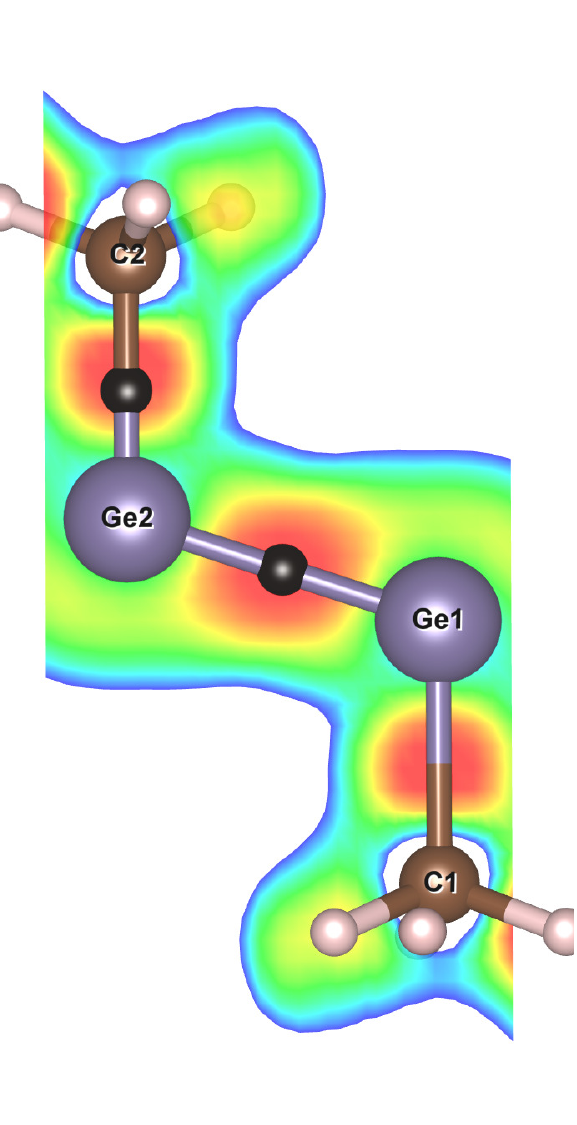}}
  \\
  \subfloat[-COOH]{\includegraphics[width=2.5cm]{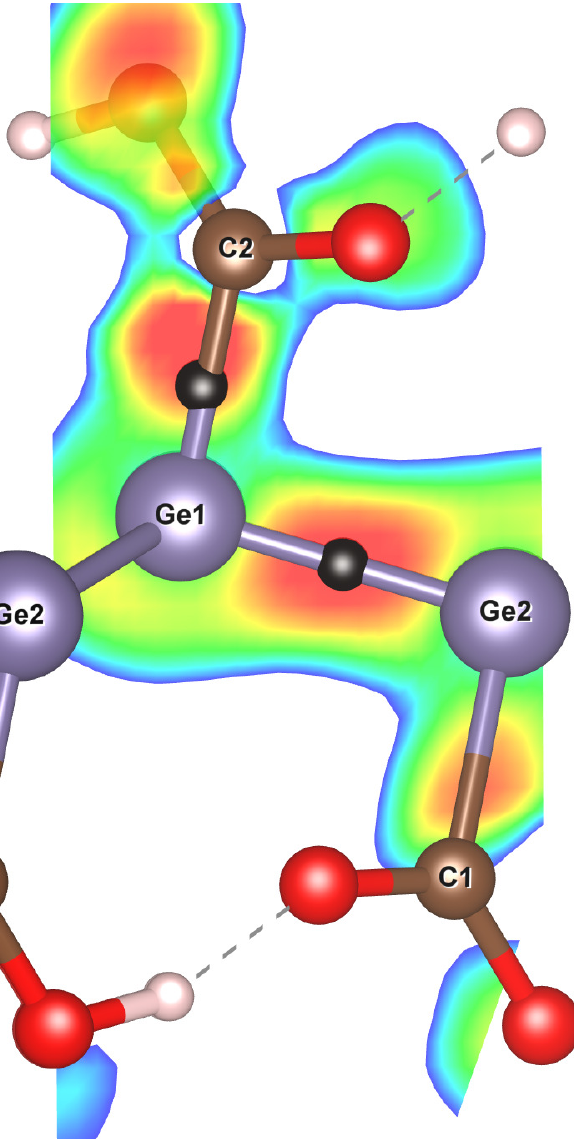}}
  \subfloat[-CH$_2$CHCH$_2$]{\includegraphics[width=3.5cm]{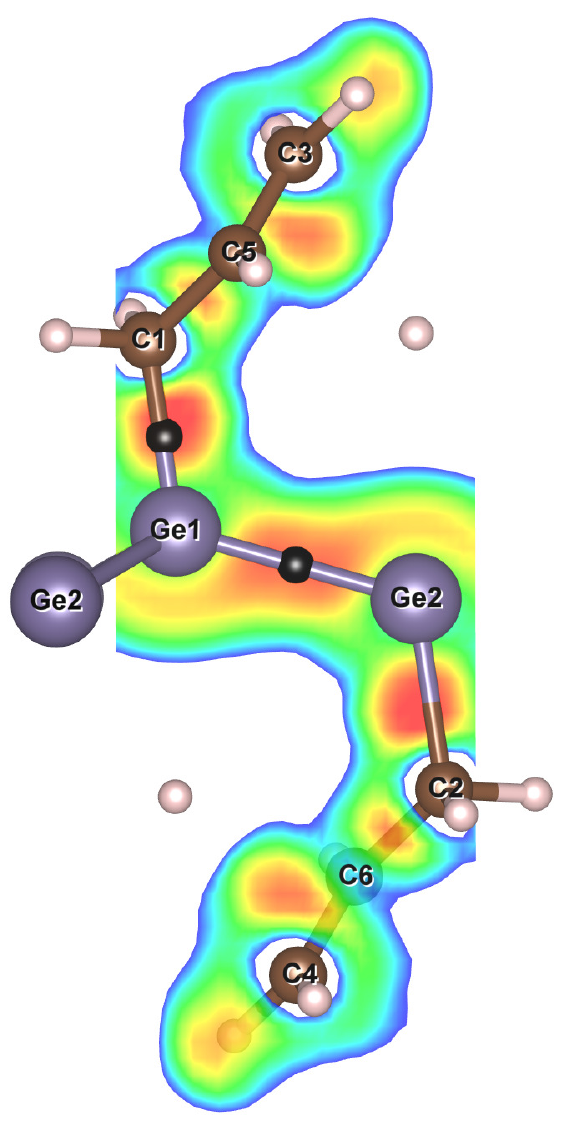}}
  \subfloat[-CH$_2$OCH$_3$]{\includegraphics[width=3.5cm]{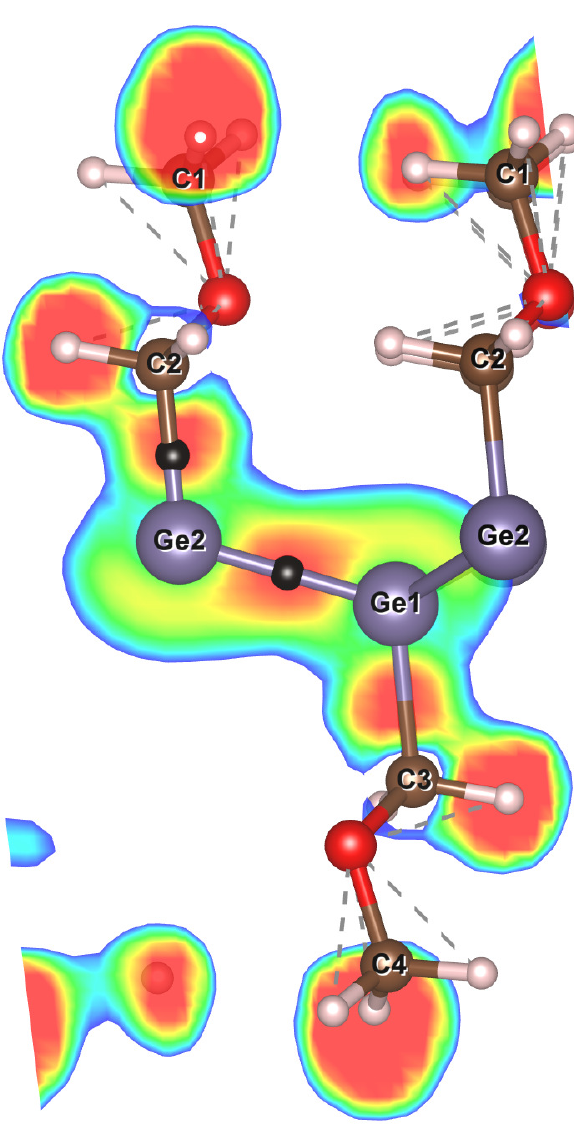}}
  \caption{Cross section of the electron localizaton function (ELF) and the main critical
  points between Ge-Ge and Ge-X bonds (black spheres) for the modified germanene
  layers. ELF values are expressed between 0.5 (blue surfaces) and 0.9 (red surfaces) .}
  \label{fig:elf}
\end{figure}

In order to extend our understanding on the chemical environment of
Ge-X layers, the electron localization function (ELF) and topological
features of the bonding critical points (BCP) were evaluated. The ELF
helps us to understand how electrons are distributed around the
atoms. ELF value close to 1 correspond to regions where there is a
high probalibility of finding electronic density characterizing
non-bonded electrons or covalent bonds.  On the other hand, ELF values
close to 0.5 indicates regions where electrons are delocalized like in
metallic bonds. As a general feature, Fig.\,\ref{fig:elf} shows a
cross section of the ELF for bare and Ge-X layers. In all systems,
there is a localized electronic density region between germanium
atoms. This region between germanium bonds is located at half distance
between germanium atoms, indicating a covalent bond type.

For buckled germanene, shown in Fig.\,\ref{fig:elf}(a), a localized
region right above and below the germanium atoms is seen. This region
is associated to the non-bonded electrons present in this system. This
comes from the fact that germanium is four-fold coordinated and in
buckled germanium one chemical bond is missing, thus leaving out one
non-bonded electron per germanium atom.

For the functionalized systems, shown in Figs.\,\ref{fig:elf} (b)-(f),
it is also seen an localized electronic density region between germamium and
hydrogen/carbon bonds. Differently from the Ge-Ge bonds, Ge-H and
Ge-C bonds are slightly asymmetric. Therefore, we can conclude that
chemical bonds between Ge-H and Ge-C are polarized, specially for the
larger organic groups.  Despite the polarization seen between Ge-X
bonds, the shared electron regions are also present between the Ge-C
distances, which is also an indication of covalent-like bonds.

\clearpage

In Fig.(\ref{fig:elf}) we see the main bonding critical points (BCP)
between Ge-Ge and Ge-X atoms. The BCP are in agreement with the ELF,
which indicates shared electrons regions between the Ge-Ge, Ge-H and
Ge-X atoms.  The interpretation of the CP give us the nature of the
chemical environment along the bonds. For light atoms in the chemical
bonds, large values of the electronic density at the BCP
($\rho_{BCP}$) and a negative value of its Laplacian ($\nabla^2
\rho_{BCP}$) are indication of shared electrons interactions
\cite{Bader1990,Jenkins2002}. However, we do not have light atoms
only. Thus, the Laplacian can assume any positive or negative value
close to zero for shared electrons regions (covalent-like
bonds)\cite{Macchi2003,Macchi1998}. The values for $\rho_{BCP}$ and
$\nabla^2 \rho_{BCP}$ are presented in Table (\ref{cpGeX}) and
(\ref{cpGeGe}).  For Ge-X bonds the $\rho_{BCP}$ presents values close
to 0.1 e\AA$^{-3}$ whereas for Ge-Ge bonds the value is smaller and
close to 0.06-0.07 e\AA$^{-3}$. For non-interacting regions with weak
interactions, such as van der Waals complexes for example $\rho_{BCP}$ is
usually much smaller than the values seen for Ge-Ge or Ge-X bonds. The
Laplacian values for Ge-Ge bonds are all negative and close to zero,
as it can be seen in Table (\ref{cpGeGe}), whereas  $\nabla^2
\rho_{BCP}$ of the Ge-X bond are close to zero and the signals
in general are positive, Table (\ref{cpGeX}) . 

\begin{table}[htbp!]
  \begin{center}
  \caption{The topological features of the Ge-X bonding critical point on the
           different chemical enviroment of buckled and Ge-ligand layers.}
  \label{cpGeX}
  \begin{tabular*}{1.0\textwidth}{@{\extracolsep{\fill}}lcccccc}
  \hline
  \hline
  Property & -H & -CH$_3$ & -COOH & -CH$_2$OCH$_3$ & -CH$_2$CHCH$_2$ \\
  \hline
  \hline
  $\rho_{BCP}$                     &  0.127 &  0.116 &  0.105 &  0.117 &  0.109 \\
  $\nabla^2 \rho_{BCP}$            & -0.066 &  0.053 &  0.013 &  0.057 & -0.007 \\
  G($\rho_{BCP}$)                  &  0.092 &  0.079 &  0.067 &  0.081 &  0.072 \\
  V($\rho_{BCP}$)                  & -0.200 & -0.145 & -0.131 & -0.147 & -0.145 \\
  H($\rho_{BCP}$)                  & -0.108 & -0.066 & -0.064 & -0.067 & -0.073 \\
  V($\rho_{BCP}$)/G($\rho_{BCP}$)  &  2.179 &  1.832 &  1.952 &  1.824 &  2.023 \\
  H($\rho_{BCP}$)/$\rho_{BCP}$     & -0.854 & -0.569 & -0.608 & -0.567 & -0.671 \\
  \hline
  \hline
  \end{tabular*}
  \end{center}
\end{table}

\clearpage

\begin{table}[htbp!]
  \begin{center}
  \caption{The topological features of the Ge-Ge bonding critical point on the
           different chemical enviroment of buckled and Ge-ligand layers.}
  \label{cpGeGe}
  \begin{tabular*}{1.0\textwidth}{@{\extracolsep{\fill}}lcccccc}
  \hline
  \hline
  Property & Bare & -H & -CH$_3$ & -COOH & -CH$_2$OCH$_3$ & -CH$_2$CHCH$_2$ \\
  \hline
  \hline
  $\rho_{BCP}$                     &  0.073 &  0.073 &  0.071 &  0.070 &  0.065 &  0.058 \\
  $\nabla^2 \rho_{BCP}$            & -0.014 & -0.023 & -0.027 & -0.030 & -0.032 & -0.024 \\
  G($\rho_{BCP}$)                  &  0.037 &  0.037 &  0.035 &  0.034 &  0.030 &  0.025 \\
  V($\rho_{BCP}$)                  & -0.077 & -0.080 & -0.076 & -0.075 & -0.068 & -0.056 \\
  H($\rho_{BCP}$)                  & -0.040 & -0.043 & -0.042 & -0.041 & -0.038 & -0.031 \\
  V($\rho_{BCP}$)/G($\rho_{BCP}$)  &  2.092 &  2.153 &  2.194 &  2.222 &  2.266 &  2.241 \\
  H($\rho_{BCP}$)/$\rho_{BCP}$     & -0.549 & -0.581 & -0.587 & -0.594 & -0.587 & -0.535 \\
  \hline
  \hline
  \end{tabular*}
  \end{center}
\end{table}

  The evaluation of $\rho_{BCP}$ and $\nabla^2 \rho_{BCP}$ is not
sufficient to characterize the nature of the chemical bond of more
complex system like Ge-X layers. Energetic features as kinetic energy
density G($\rho_{BCP}$), the potential energy density
V($\rho_{BCP}$) and the total energy density H($\rho_{BCP}$) can
also be helpful for interpreting the nature of the chemical
bond. Bianchi \emph{et al.} \cite{Bianchi2000} suggested that
G($\rho_{BCP}$) $\ll$ $\vert$ V($\rho_{BCP}$) $\vert$,
V($\rho_{BCP}$) $\ll$ 0 and H($\rho_{BCP}$) $\ll$ 0 are indicative of
the presence of the covalent bonds.  For the BCP between Ge-X,
G($\rho_{BCP}$) is much smaller than the
$\vert$V($\rho_{BCP}$)$\vert$, as seen in Tab\,\ref{cpGeX}. However,
for Ge-Ge bonds, G($\rho_{BCP}$) is smaller and closer to
$\vert$V($\rho_{BCP}$)$\vert$, Tab\,\ref{cpGeGe}. The evaluated values for
V($\rho_{BCP}$) and H($\rho_{BCP}$) are negative for both BCP at
Ge-Ge and Ge-X bonds, Tab\,\ref{cpGeX} and Tab\,\ref{cpGeGe}. 
These energetic conditions indicate shared electrons interactions, i.e., 
covalent bonds between Ge-Ge and Ge-X.

  The ratios $\vert$V($\rho_{BCP}$)$\vert$/G($\rho_{BCP}$) and
H($\rho_{BCP}$)/$\rho_{BCP}$ are other topological quantities which
are used to address the nature of the chemical bond environment
\cite{Espinosa2002,Macchi2003,Macchi1998}. For covalent bonds, it is
common to obtain the ratio
$\vert$V($\rho_{BCP}$)$\vert$/G($\rho_{BCP}$) $>$ 2, in intermediate
bond cases such as high polarized and ionic interactions it is found
${\rm 1 < \vert V(\rho_{BCP}) \vert /G(\rho_{BCP}) < 2}$.  Table \ref{cpGeGe}
shows ${\rm \vert V(\rho_{BCP})\vert/G(\rho_{BCP})> 2}$ for all Ge-X layers
critical points between Ge-Ge indicating the presence of a covalent
bonds. For the critical points localized between Ge-X atoms, the values
are smaller and/or closer to 2.  This indicates polarized
covalent bonds between Ge-C atoms, as it can be seen in Table
\ref{cpGeX}. The other ratio H($\rho_{BCP}$)/$\rho_{BCP}$ feature is
smaller than zero for all localized critical points which is also an
indicative of covalence between the Ge-Ge and Ge-X
atoms\,\cite{Espinosa2002,Macchi2003,Macchi1998}.

\subsection{Electronic properties}

\begin{figure}[htpb!]

\includegraphics[width = 7.5cm, scale=1, clip = true, keepaspectratio]{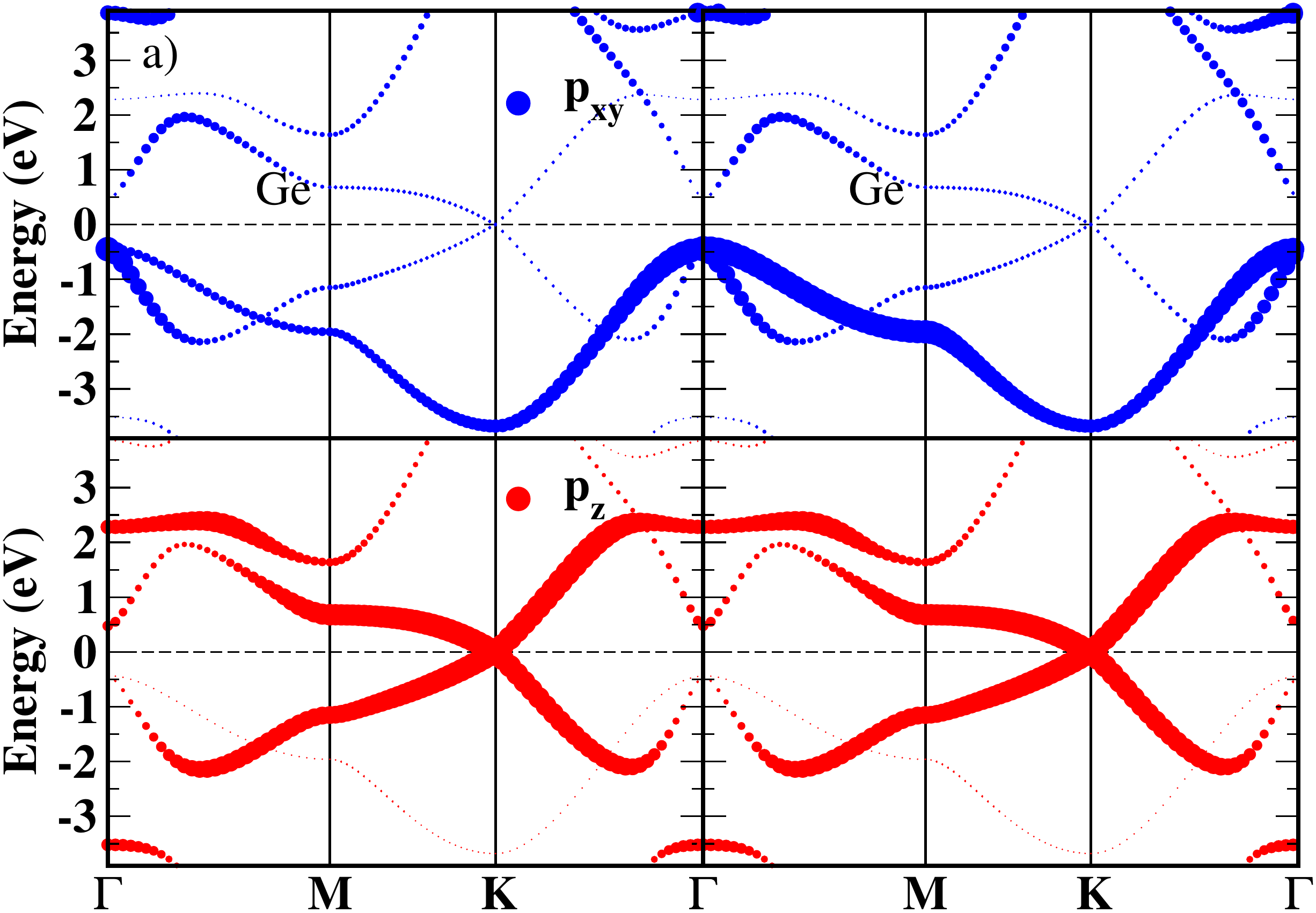}
\includegraphics[width = 7.5cm, scale=1, clip = true, keepaspectratio]{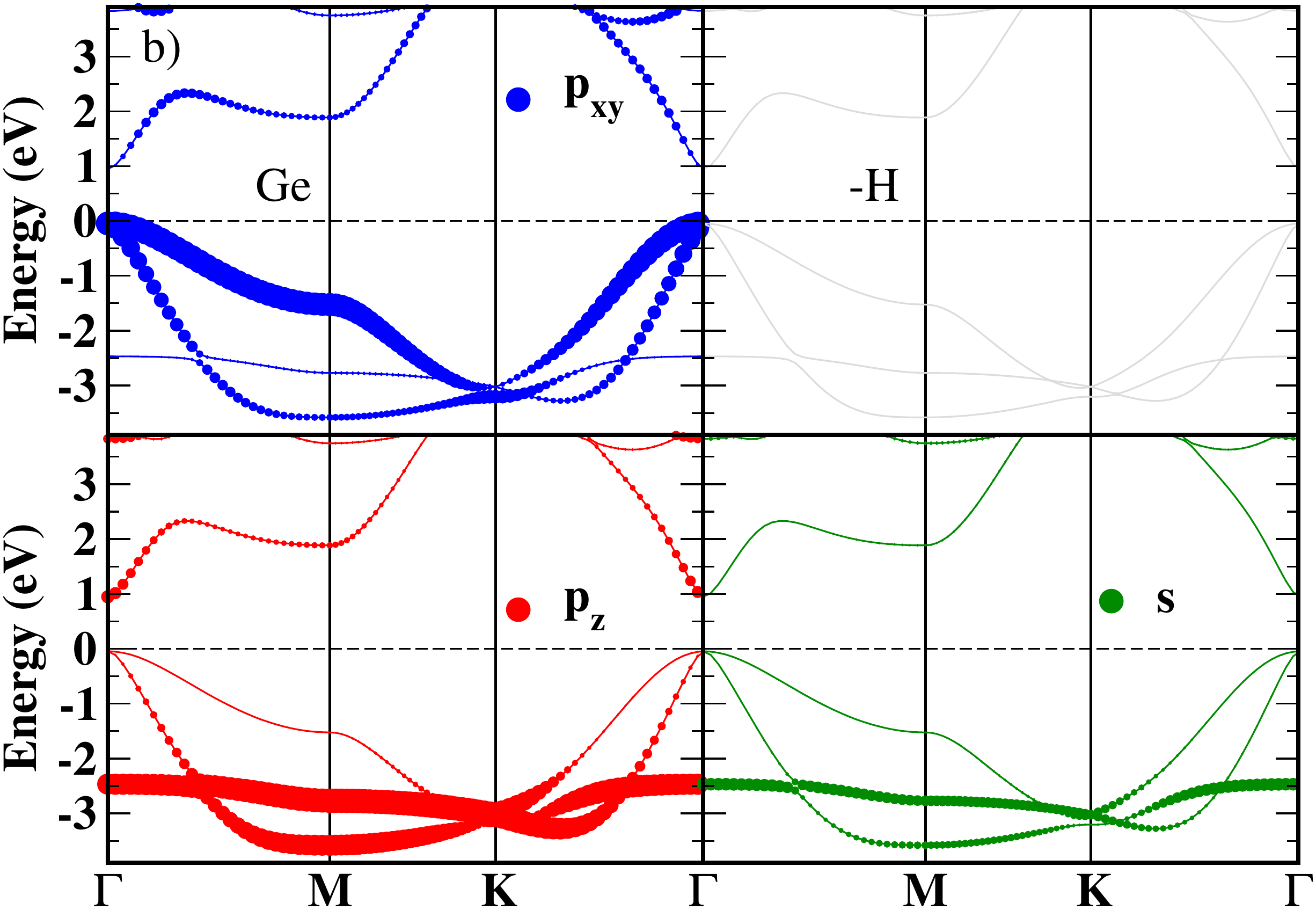}
\includegraphics[width = 7.5cm, scale=1, clip = true, keepaspectratio]{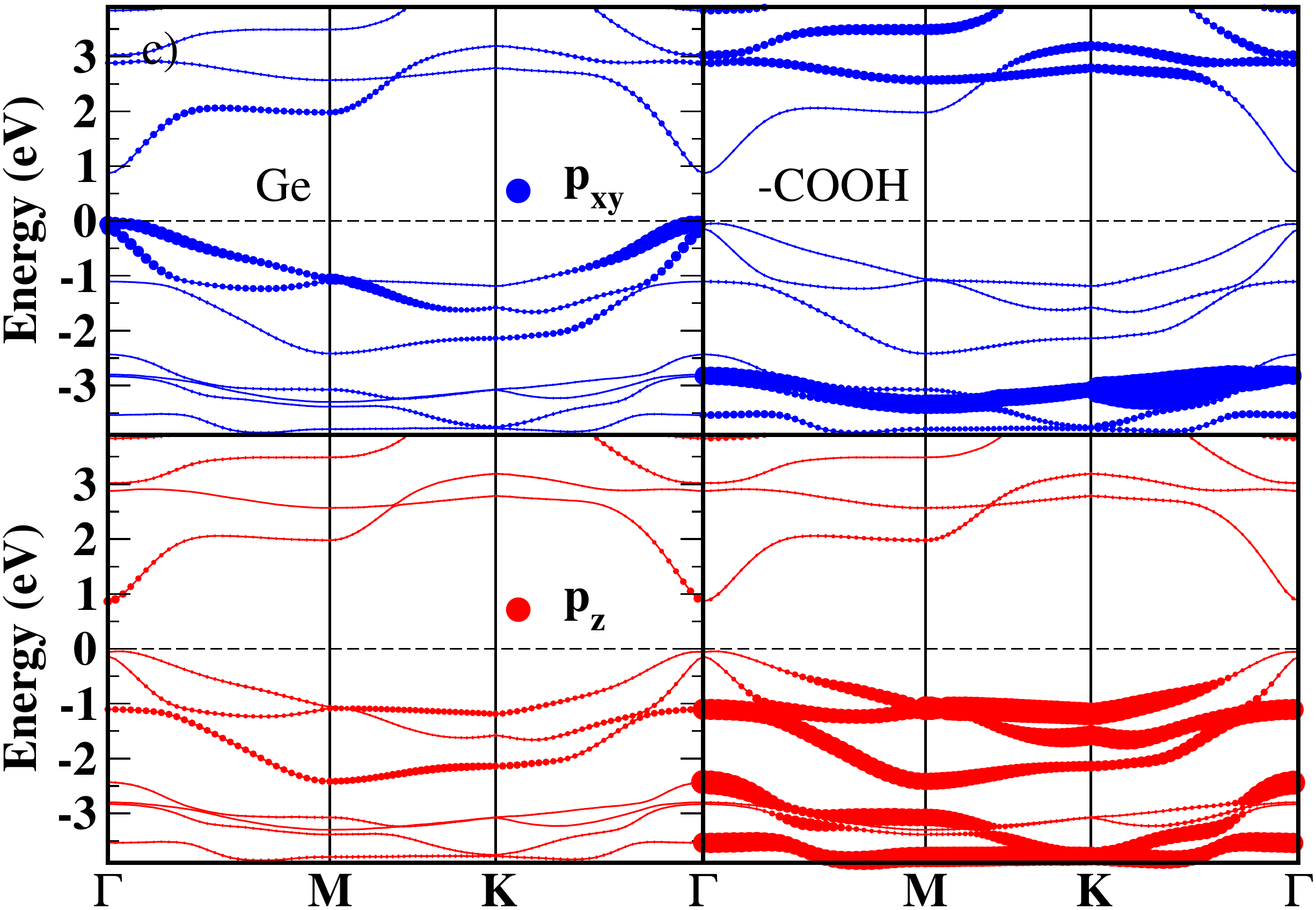}
\includegraphics[width = 7.5cm, scale=1, clip = true, keepaspectratio]{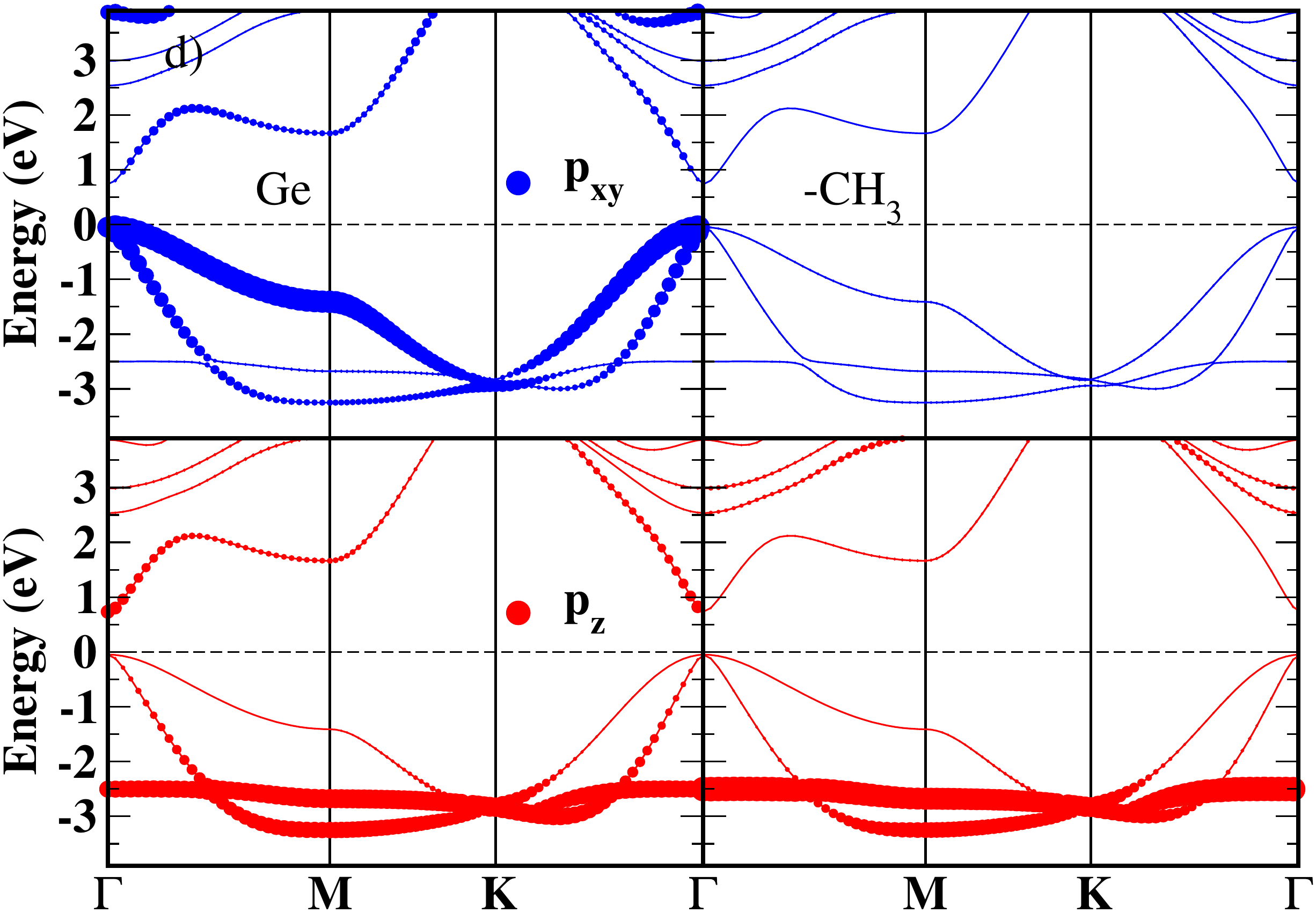}
\includegraphics[width = 7.5cm, scale=1, clip = true, keepaspectratio]{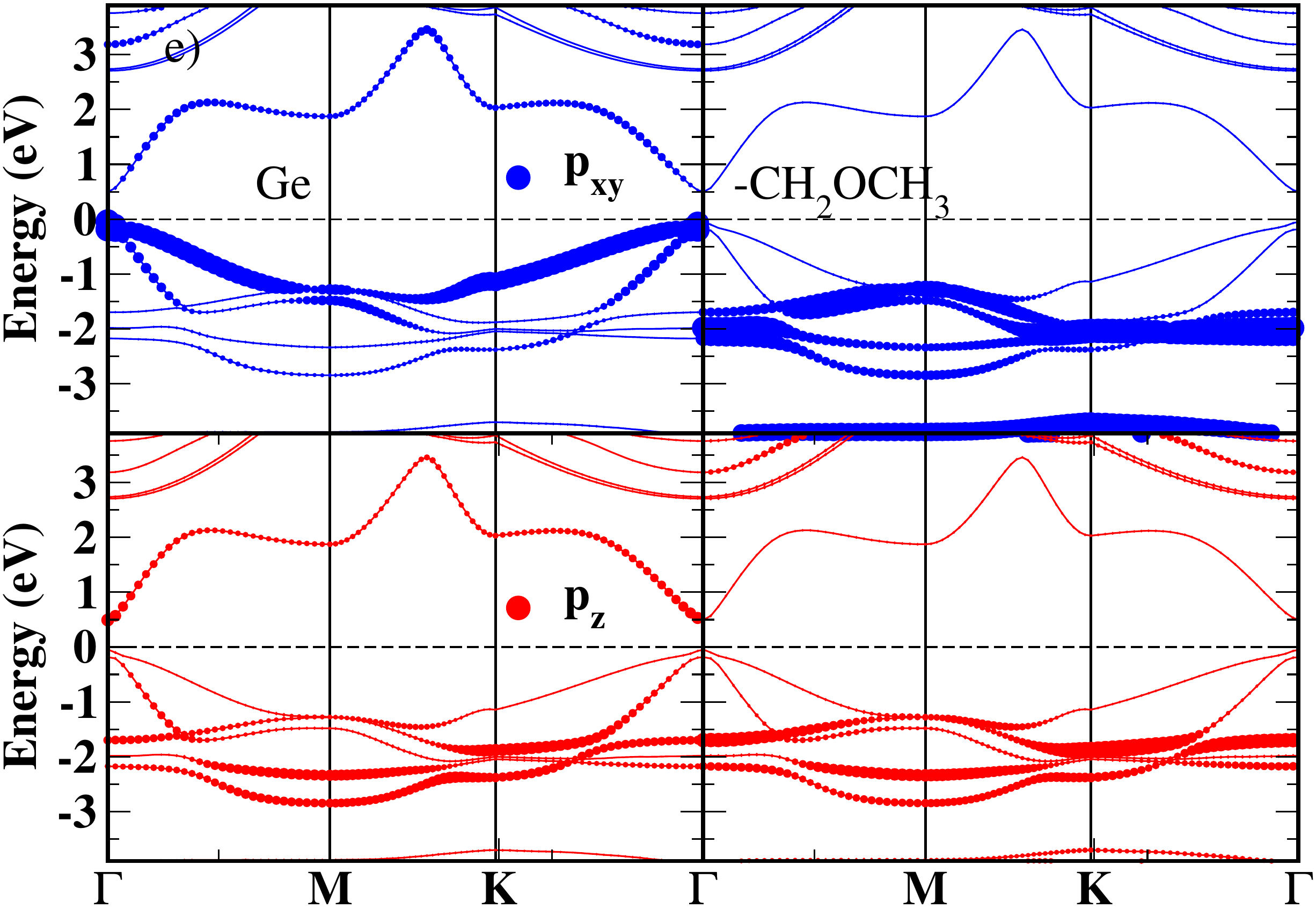}
\includegraphics[width = 7.5cm, scale=1, clip = true, keepaspectratio]{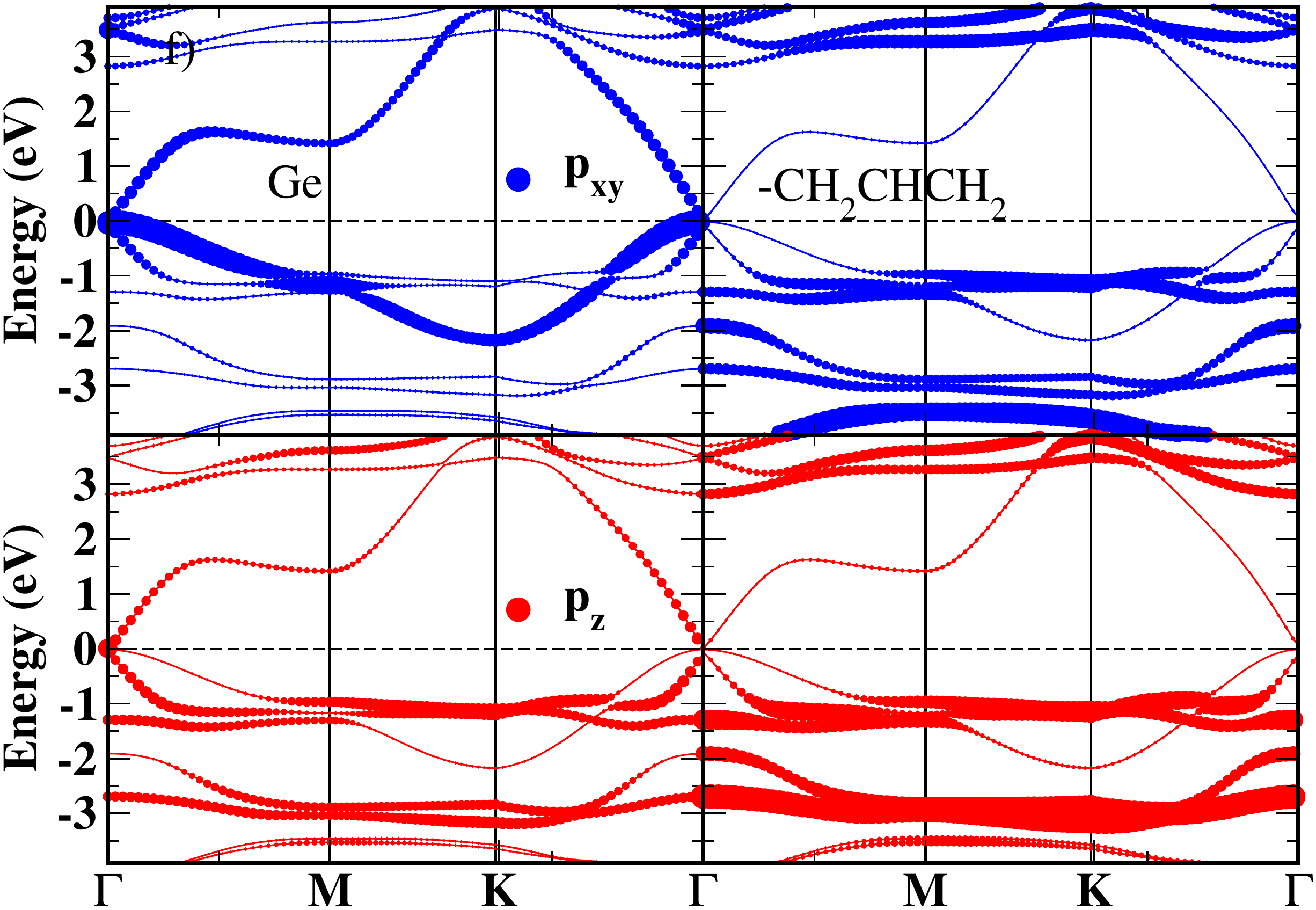}
\caption{Orbital resolved band structure of modified germanene structures. a) germanene, b) germanane, c) -COOH, d) -CH$_3$, e) -CH$_2$OCH$_3$ and f) CH$_2$CHCH$_2$ . The zero of energy is set to the VBM. }
\label{fig:projected_bands}
\end{figure}

In order to understand the interaction between ligand and the
substrate, we have calculated the individual orbital contributions to
the band structure, as shown in Fig.\,\ref{fig:projected_bands}. The
p$_{\rm xy}$, p$_{\rm z}$ and s contributions are shown. Bare
germanium layers have a metallic character and zero gap with a linear
dispersion at the $\Gamma$ point, as it is seen in
Fig.\ref{fig:projected_bands} (a). Band-to-band $\Gamma-\Gamma$ and
$\Gamma$-M energy transitions are reported in
Table\,\ref{table:electronic}.

As the size of ligand increases, the band gap decreases. It is
smaller upon adsorption of -H and larger upon adsorption of
${\rm  -CH_2OCH_3}$.  The band structure for Ge-H is shown in
Fig.\,\ref{fig:projected_bands}(b), for ${\rm Ge-CH_3}$(c) in
Fig.\,\ref{fig:projected_bands}(c), for Ge-COOH in
Fig.\,\ref{fig:projected_bands}(d) and for ${\rm Ge-CH_2OCH_3}$ in
Fig.\,\ref{fig:projected_bands}(e).

Contributions to the states close to the Dirac point
are mainly due to $p_{\rm xy} = p_{\rm x} + p_{\rm y}$ and $p_{\rm z}$ orbitals. Upon ligand
adsorption on germanium layers, the Dirac cone moves from K to
$\Gamma$ point. 
All these hybrid sttructure have
now a sizeable band gap. The exception is the ${\rm Ge-CH_2CHCH_2}$,
which shows metallic character, as shown in
Fig.\,\ref{fig:projected_bands}(f). This can be understood considering
that this group is an electron donor and therefore renders a metallic
system.

${\rm -CH_2OCH_3}$ induces electron capture and induces band gap
opening. This can be explained by noting the ${\rm -CH_2OCH_3}$ group is an electron
acceptor. This means that the ligand character is also important to
determine the bond strength and consequently the electronic structure
of the hybrid system.

We can clearly see contributions from the functional groups at VBM and
CBM, implying that we have formation of bonds between the ligand and
the germanium layers. The change in the band gap is in agreement with
suggestions of Ref.\,\cite{JiangNT:2014}. Our values are somewhat
underestimated compared to experimental ones.

The dielectric function of hybrid germanene-organics layers were
investigated by calculating the imaginary part of the dielectric
function at GW level. The imaginary part of the dielectric function is
calculated directly from the electronic structure through the joint
density of states and the momentum matrix elements occupied and
unoccupied eigenstates according Ref.\cite{Shishkin:07}.

\begin{figure}[htbp!]
\includegraphics[width = 7cm,scale=1, clip = true, keepaspectratio]{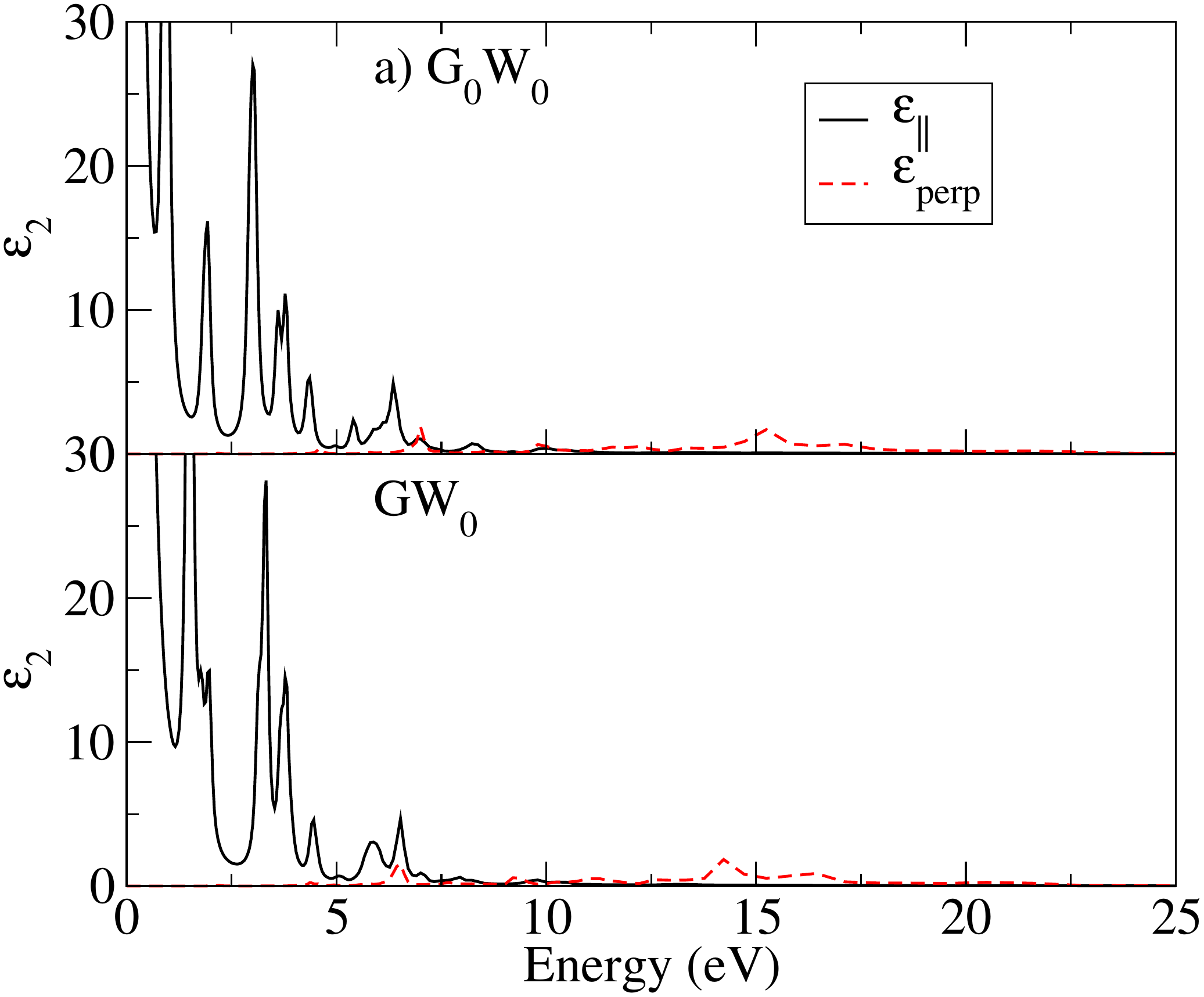}
\includegraphics[width = 7cm,scale=1, clip = true, keepaspectratio]{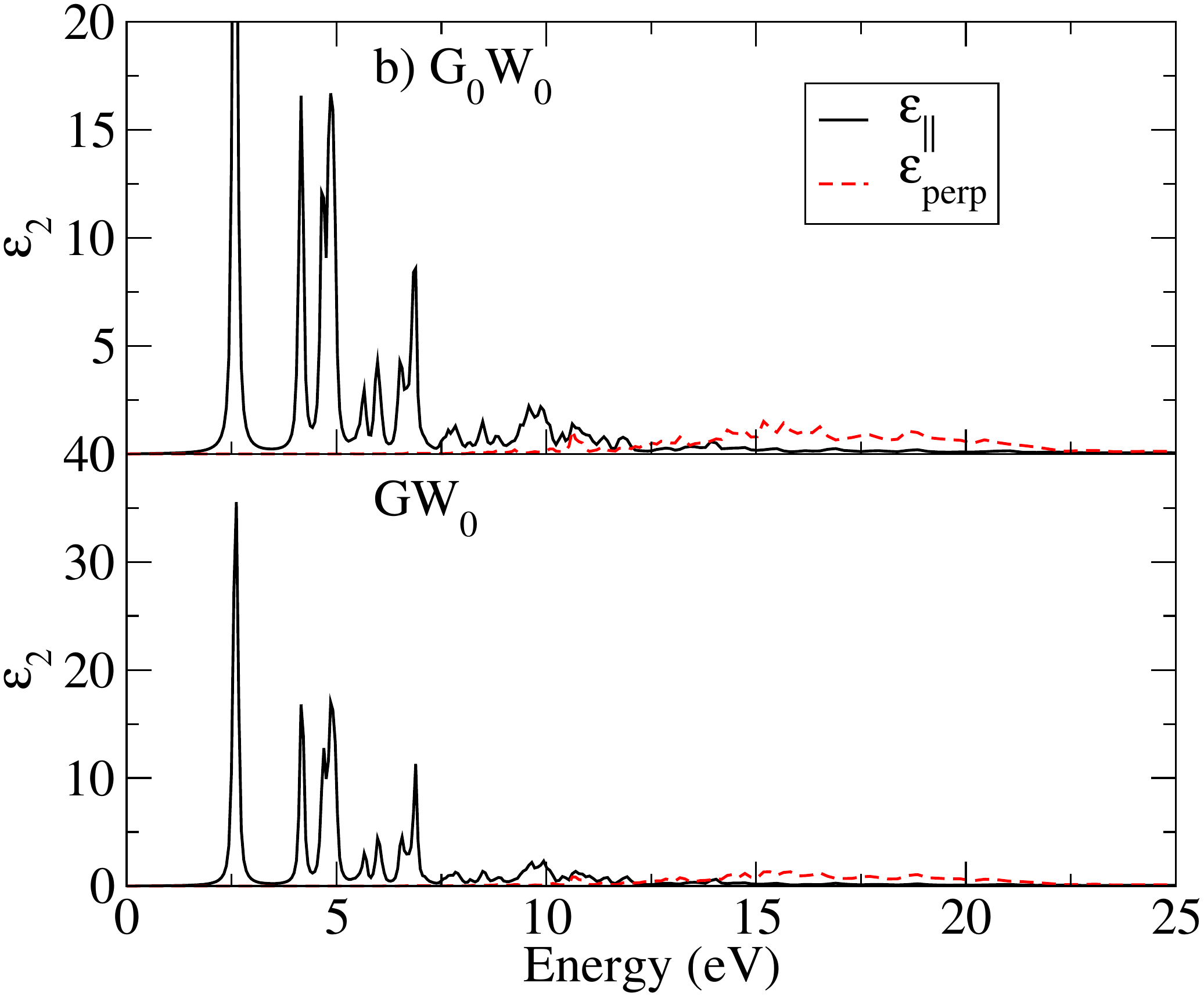}
\includegraphics[width = 7cm,scale=1, clip = true, keepaspectratio]{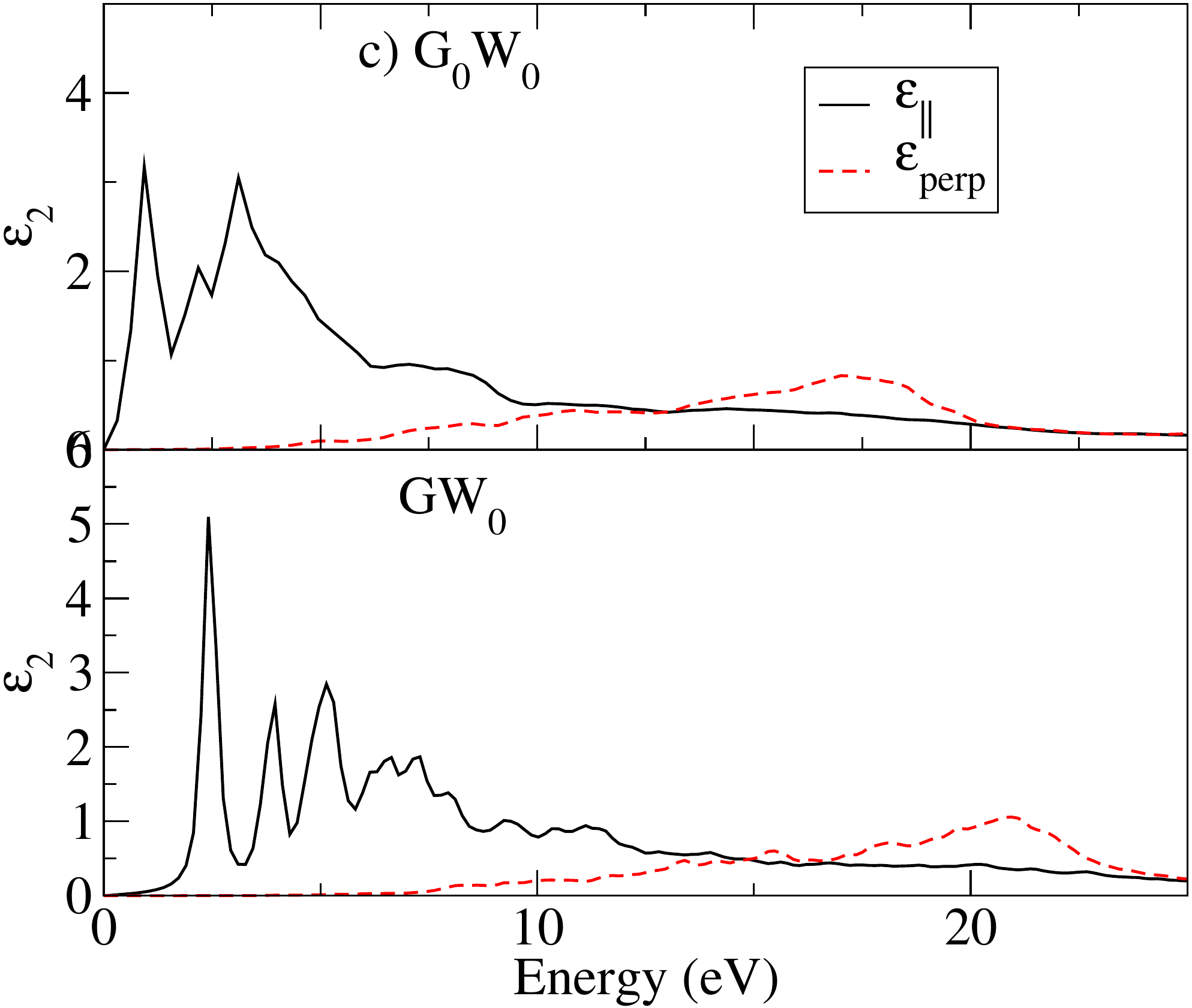}
\includegraphics[width = 7cm,scale=1, clip = true, keepaspectratio]{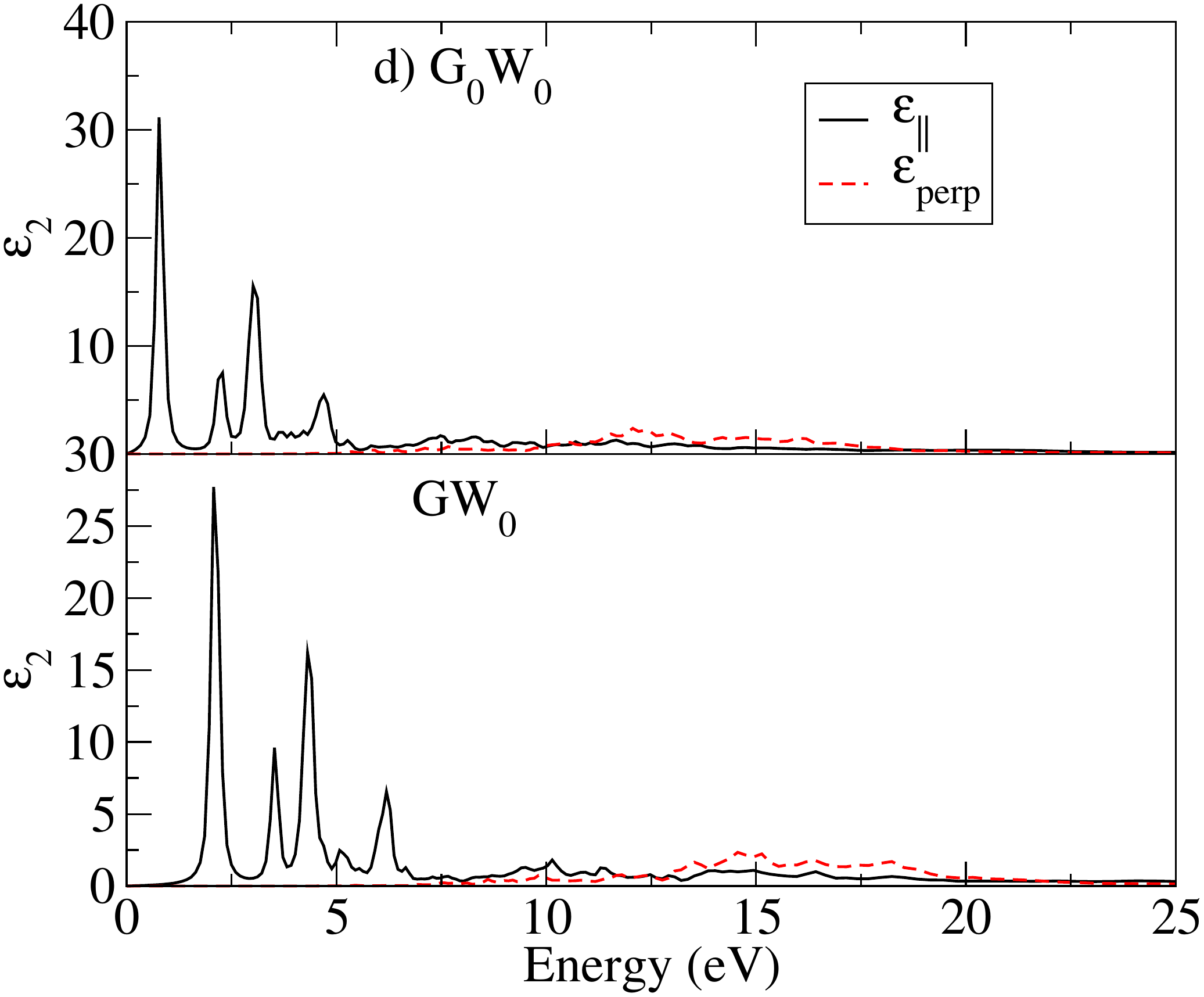}
\includegraphics[width = 7cm,scale=1, clip = true, keepaspectratio]{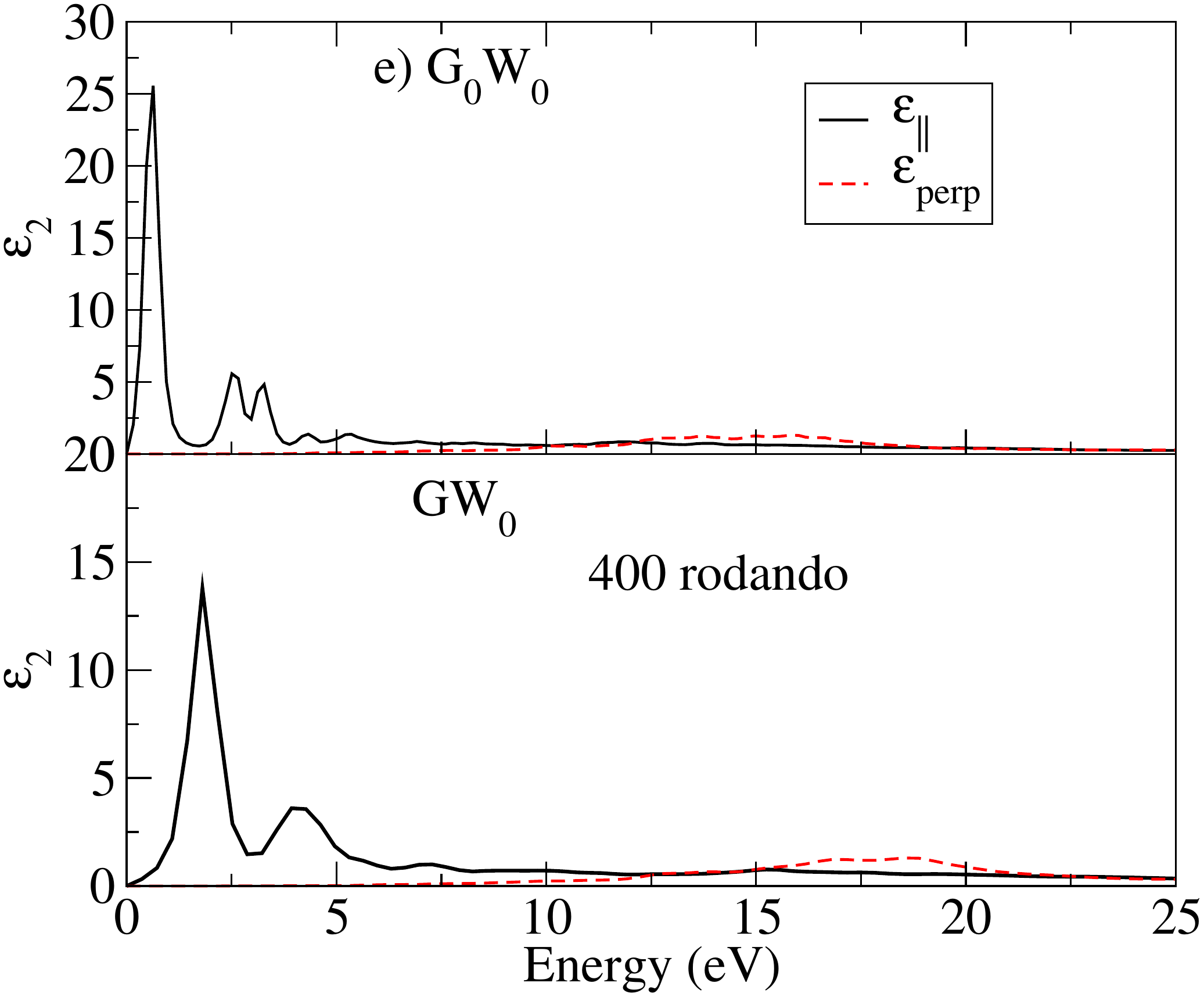}
\includegraphics[width = 7cm,scale=1, clip = true, keepaspectratio]{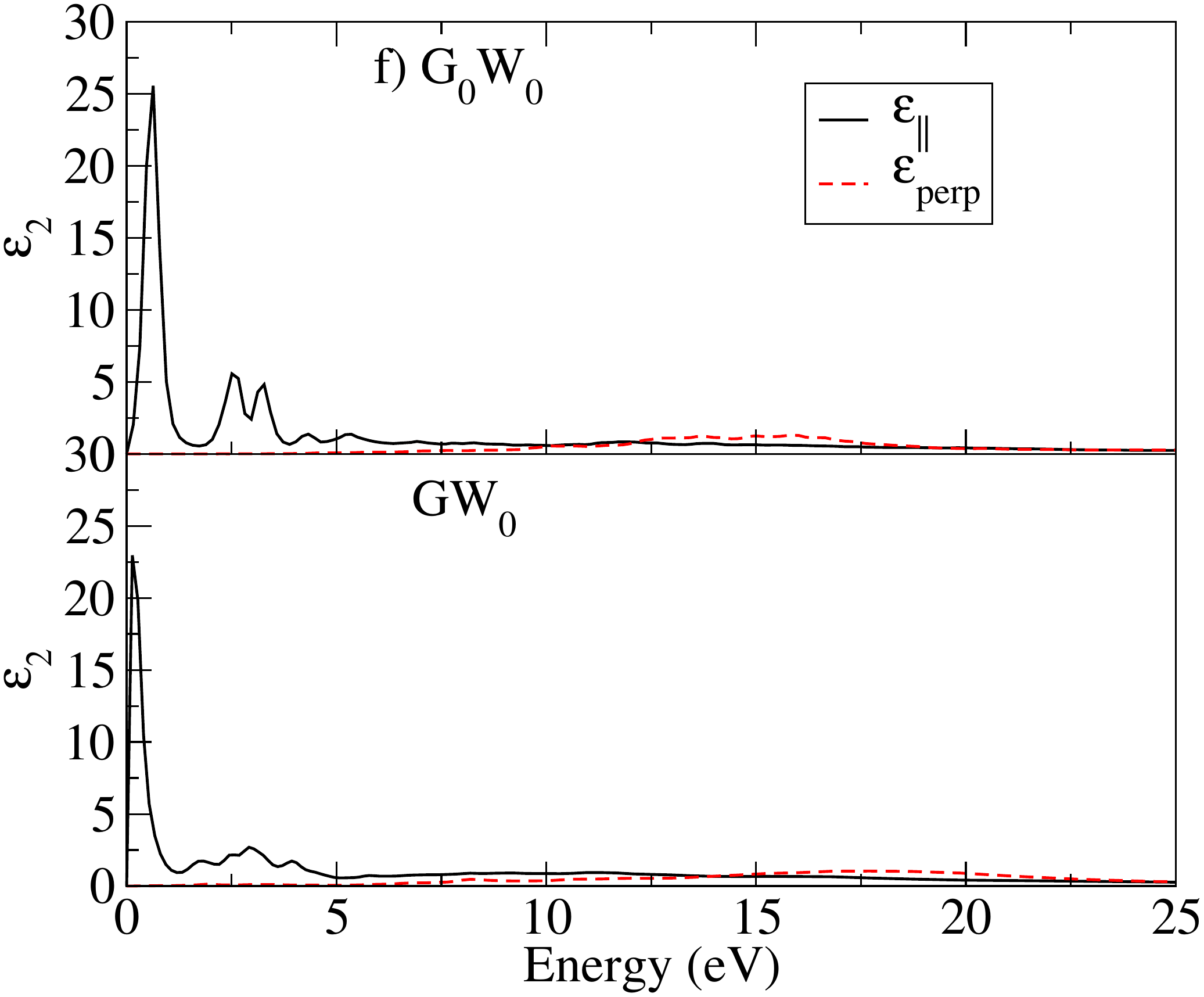}
\caption{\label{fig:diel_GW} Imaginary part of the dielectric function for pure and 
  functionalized germanene within G$_0$W$_0$ (upper panel on each figure) and GW$_0$ 
  (lower panel on each figure) approximations. a) germanene, b) germanane, c) -COOH, d) -CH$_3$, 
  e) -CH$_2$OCH$_3$ and f) CH$_2$CHCH$_2$.  Light propagation parallel to the germanium layer 
  is denoted as $\epsilon_{||}=(\epsilon_{xx}+\epsilon_{yy})/2$. Light propagation perpendicular to 
  the germanium layers is denoted as $\epsilon_{\rm perp}=\epsilon_{zz}$.}
\end{figure}

We show the dielectric function calculated within the G$_0$W$_0$ and
GW approximations in Fig.\,\ref{fig:diel_GW}. The averaged parallel
$\varepsilon_{\|} = (\varepsilon_{\rm xx}+\varepsilon_{\rm yy})/2$ and
perpendicular $\varepsilon_{\perp} = \varepsilon_{zz}$ components of
the the imaginary part of the dielectric function $\varepsilon_2$ are
shown. The $\varepsilon_{\|}$ component corresponds to the propagation
of the external electromagnetic field parallel to the germanene plane
while $\varepsilon_{\perp}$ corresponds to the field perpendicular to
the plane. Because of optical selection rules, anisotropy in the optical spectra
is seen. Anisotropy has also been reported in layered monochalcogenide
of germanium sulfide (GeS) \cite{GeS}, black phosphorous\,\cite{blackP} and bismuthene\cite{Ciraci2019}.

The systems with a gap show finite absorption limits for both parallel and perpendicular
directions with larger intensity for the $(\varepsilon_{\|}$
component.

\begin{table}[ht!]
  \caption{\label{table:electronic} Energy gaps and transitions (in eV) corresponding to Fig.\,\ref{fig:diel_GW}) of ligand modified germanene calculated within GGA, G$_0$W$_0$ and GW.}
  \begin{center}
\begin{tabular*}{1.0\textwidth}{@{\extracolsep{\fill}}lcccccccc}
  \hline
  \hline
  ligand        &  \multicolumn{3}{c}{gap} &   \multicolumn{2}{c}1$^{st}$ peak   &   \multicolumn{2}{c}2$^{nd}$ peak & Exp.\,\cite{JiangNT:2014} \\
  \hline\hline
                & PBE  & G$_0$W$_0$ & GW & G$_0$W$_0$ & GW & G$_0$W$_0$ & GW   &       \\
  \hline
bare (metal)           &  &  &   &     &     &     &     \\
-H              & 0.99 & 2.1 &  2.5 & 2.5 & 2.6 & 4.1 & 4.2 & 1.57\\
-COOH           & 0.91 & 2.0 &  2.4 & 0.9 & 2.5 & 2.1 & 4.0 & \\
-CH$_3$         & 0.78 & 1.7 &  2.1 & 0.8 & 2.0 & 2.4 & 3.6 & 1.66\\
-CH$_2$OCH$_3$  & 0.54 & 1.6 &  1.6 & 0.6 & 1.4 & 2.6 & 3.3 & 1.45\\
-CH$_2$CHCH$_2$ (metal) &      &     &      &     &     &     &     \\
\hline
\hline
\end{tabular*}
  \end{center}
\end{table}

As discussed in Ref.\cite{JiangNT:2014}, both ligand size and
electronegativity can change the bond length and band gap of
functionalized germanene.  Larger ligands are expected to lead to
larger Ge-Ge separation, thus yielding a lower band gap. Ligands with
greater electronegativity are expected to withdraw electrons and
therefore lower the band gap. The size of the ligands we have
calculated decreases in the order -CH$_2$CHCH$_2$ $<$ -CH$_2$OCH$_3$ $<$
-CH$_3$ $<$ –H.  On the other hand ligand electronegativity decreases
in the order from -CH$_2$OCH$_3$ $<$ -H $<$ -CH$_3$ $<$
-CH$_2$CHCH$_2$. According to these experiments, the band gap shoud be
inversely proportional to Ge-Ge bond length. The observed band gap
value increases with decreasing ligand electronegativity, with the
exception of Ge-CH$_2$CHCH$_2$. The general conclusion is that ligands
that are more electron-withdrawing and have greater steric bulk will
expand the Ge-Ge framework and lower the band gap, with complete
ligand coverage.

We find that the inclusion of self-consistency in the Green's
functions leads to a blue shift compared to the G$_0$W$_0$
results. The bare germanene shown in Fig. \ref{fig:diel_GW}(a) has
metallic behavior.  G$_0$W$_0$ increases the gap and GW$_0$ increases
further. As a general feature, two main absorption peaks appear in the
spectrum. Ge-H shown in Fig.\,\ref{fig:diel_GW}(b) has a peak at
2.5\,eV at G$_0$W$_0$ and 2.6 eV at GW$_0$. A second peak appears at
4.1 and 4.2 eV, at G$_0$W$_0$ and GW$_0$ levels, respectively. The
Ge-COOH spectrum shown in Fig.\,\ref{fig:diel_GW}(c) has peaks at 0.9
and 2.5 eV. A second peak appears at 2.1 and 4.0 at  G$_0$W$_0$ and GW$_0$ levels, respectively.
A first peak at 0.8 and 2.0 for Ge-CH$_3$ are seen in
Fig.\,\ref{fig:diel_GW}(d) at G$_0$W$_0$ and GW$_0$ levels,
respectively. Another peak at 2.4 and 3.6 eV appears at G$_0$W$_0$ and
GW$_0$ levels, respectively. Last, ${\rm Ge-CH_2OCH_3}$ shows high
intensity at at 0.6 and 1.4 eV, at G$_0$W$_0$ and GW$_0$ levels, as
it can be seen in Fig.\,\ref{fig:diel_GW}(e). On the other hand
another peak appears at 2.6 and 3.3 eV G$_0$W$_0$ and GW$_0$ levels,
Finally, we find that ${\rm -CH_2CHCH_2}$ is an electron donor and
therefore renders the Ge-${\rm -CH_2CHCH_2}$ metallic.

According to Ref.\,\cite{JiangNT:2014} the absorption onset measured
via Diffuse reflectance absorption (DRA) for germanium modified layers
is 1.57\,eV for Ge-H, 1.66\,eV for GeCH$_3$, 1.55 eV for
GeCH$_2$CHCH$_2$ and 1.45\,eV for Ge-CH$_2$OCH$_3$. Our GW$_0$
calculations are in reasonable agreement with these experimental
results, since the band gap decreases with the ligand size. The
excpetion is for Ge-CH$_2$CHCH$_2$, where the experimental value for
the band gap is 1.55\,eV, but we found this materials to be a metal.
The reason for such discrepancy could be inferred taken into account
the number of layers, coverage and adsorption site. However, further
investigation is needed to clarify this aspect.

\section{Conclusions}

We have performed first-principles calculations of germanene
functionalized layers with small organic ligands. Our charge density
analysis show that the ligands are chemisorbed on the germanium
layers. Our calculations for the dielectric properties of bare and
ligand adsorbed germanene show a large anisotropy and that the
absorption onset is determined by both ligand electronegativity and
size, in reasonable agreement with recent experimental
results\,\cite{JiangNT:2014}. We believe our findings of a finite gap shows open a
path for rational design of nanostructures with possible applications
in biosensors and solar cells.

\section{Acknowledgements}

We acknowledge the financial support from the Brazilian Agency CNPq
and German Science Foundation (DFG) under the program FOR1616. The
calculations have been performed using the computational facilities of
Supercomputer Santos Dumont and at QM3 cluster at the Bremen Center
for Computational Materials Science and CENAPAD.

%\bibliography{references}
\bibliographystyle{apsrev}

\end{document}